\documentclass[useAMS,usenatbib]{mn2e}
\usepackage{epsfig}
\usepackage{longtable,lscape}
\usepackage{graphicx}
\usepackage{natbib}
\begin{document}


\title[Statistics on local H\textsc{II} galaxies.]
{The impact on the general properties of H\textsc{II} galaxies in the local universe of the visibility of the 
[O\textsc{III}]$\lambda 4363$ line.}
\author[C. Hoyos and A. I. D\'{\i}az]
{C. Hoyos $^{1}$\thanks{E-mail:
carlos.hoyos@uam.es} and A. I. D\'{\i}az $^{1}$\thanks{E-mail: angeles.diaz@uam.es}\\
$^{1}$Departamento de F\'{\i}sica Te\'orica, Universidad Aut\'onoma de Madrid,
Carretera de Colmenar Viejo km 15.600, Madrid 28049, Spain}

\date{\textbf{To appear in Monthly Notices of the Royal Astronomical Society.}}
\pagerange{\pageref{firstpage}--\pageref{lastpage}} \pubyear{2005}
\maketitle
\label{firstpage}

\begin{abstract}
We present a statistical study of a very large sample of H\textsc{II}
galaxies taken from the literature. We focus on the differences in several
properties between galaxies that
show the auroral line [O\textsc{III}]$\lambda 4363$ and those that do not
present this feature in their spectra. It turns out that objects
without this auroral line are more luminous, more metal-rich and present a lower
ionization degree. The underlying population is found to be much more
important for objects without the [O\textsc{III}]$\lambda 4363$ line, and 
the effective temperature of the ionizing star clusters
of galaxies not showing the auroral line is probably lower.
We also study the subsample of H\textsc{II} galaxies whose properties most
closely resemble the properties of the intermediate-redshift population of 
Luminous Compact Blue Galaxies (LCBGs). The objects from this subsample are
more similar to the objects not showing the [O\textsc{III}]$\lambda 4363$
line. It might therefore be expected that the intermediate-\textit{redshift}
population of LCBGs is powered by very massive, yet somewhat aged star
clusters. The oxygen abundance of LCBGs would be greater than the average 
oxygen abundance of local H\textsc{II} galaxies.

\end{abstract}


\section{Introduction}

The term ``H\textsc{II galaxy}'' currently denotes dwarf emission line galaxies
undergoing violent star formation (VSF)\citep{meln85}, a process by which thousands
of massive stars (m$\geq$20 M$_{\odot}$) have recently been formed in a very
small volume (a few parsec in diameter) and on the timescale of only a
few million years.
HII galaxies comprise a subset of the larger class
of objects referred to as ``blue compact galaxies'' (BCG). At optical
wavelengths, the observable properties of HII galaxies are dominated by
the young component of their stellar population and their spectra are
essentially identical to those of Giant Extragalactic HII Regions
(GEHR) in nearby spiral and irregular galaxies. The analysis of these
spectra shows that many HII galaxies are metal poor objects, some of
them -- IZw18, UGC4483 -- being the most metal poor systems known.

Although it was initially thought that HII galaxies are compact,
essentially spheroidal, in fact they show a variety of morphologies
with an appreciable number of them having two or more components and
showing clear signs of interactions \citep{etelles97}. Actually, IIZw40, one 
of the first HII galaxies identified, when imaged at sufficiently 
high resolution looks like the result of a merger of two separate 
subsystems (\citet{meln92}).

The study of blue, compact star forming systems in the distant universe is an 
important ingredient in galaxy formation and evolution theories. Although such 
systems probably were not much more powerful than many local H\textsc{II}
galaxies, it is believed
that they were much more common in the past than they are
today and therefore they harboured a great amount of
the star formation density in the universe. This fact makes of high
redshift blue compact galaxies very interesting targets for
observation. Unfortunately, since they are located at large distances,
their study can only be carried out with a combination of Hubble Space 
Telescope providing high spatial resolution and 10-m 
class telescopes providing high collecting power. Nevertheless, data on
this kind of objects are accumulating fast.


Luminous Compact Blue Galaxies (LCBG) are defined as very luminous (M$_B$ $\leq$ -17.5),
compact ($\mu_{B}\leq21.5$\textrm{mag arcsec}$^{-2}$) and blue (B-V) $\leq$ 0.6) galaxies undergoing a major burst of star formation
(Hoyos et al. 2004). According to this definition LCBG, as a family, include the most luminous local
HII galaxies. In principle, since they belong to the general category of emission
line objects, the techniques used for their analysis are those already developed
for the same class of objects in the local universe. However, in order
to interpret these analyses properly in terms of evolution, a good
comparison sample needs to be available.

There have been many studies of HII galaxies in the local universe, and the errors and
uncertainities in the determination of their properties are supposed to be well understood. However, due to the
importance of the metallicity effects in controlling the gaseous emission line spectra,
most of these studies have been restricted to a subsample that allows a good direct determination of their chemical
abundances, through the detection and measure of the weak [OIII] $\lambda$
4363 \AA{} line. This implies the selection of the highest excitation objects which, in principle, can't be considered
to be a good comparison sample.

The purpose of the present work is to analyse a large sample of local universe HII galaxies. In particular two issues are
addressed. The first one is the comparison of several observational parameter
distributions for objects with and without measurements of the
[O\textsc{III}]$\lambda 4363$ \AA{}
emission line and the statistical analysis of the whole HII galaxy sample to define their average properties.
The second one is the statistical study of a subsample consisting of the most luminous
objects, local representatives of LCBG, and its
comparison to the whole HII galaxy sample.


The cosmology assumed through this paper is a flat universe with H$_{0}$=70\textrm{km s}$^{-1}$\textrm{ Mpc}$^{-1}$.

\section{The sample selection.}

We have compiled published emission line measurements of local H\textsc{II} galaxies from different 
sources. The data gathered, together with the works we have used to compile the galaxy sample 
are given in table \ref{m1y2pres}.
The vast majority of the objects selected for this study were discovered 
using Schmidt telescopes, searching for strong emission lines or blue colors.
Sources with strong emission lines are easiliy detected through objective 
prism surveys. This technique is best suited to detect objects with high
equivalent widths and line strengths. Objects with strong continua are 
lost in these surveys, since the emission lines are swallowed by the continuum.
On the other hand, galaxies with weak lines which have evolved past their peaks of 
star formation but which are still quite blue are found through colour selection techniques.

The objects from references 6 and 10 come from the Tololo \citep{smith76}
and University of Michigan (UM) \citep{macalpine77} objective prism surveys. The limiting magnitude is about 19.0.
The spectroscopic observations for these samples were taken using several telescopes and detectors, and
the observing conditions were not always good. Some of the observations were carried out at 
Las Campanas 2.5m telescope, using narrow apertures. These observations therefore can't provide absolute 
fluxes. They were also affected by second order contamination.
The rest of the spectra were obtained using the 3.6m at ESO. The slit aperture 
was 8\arcsec , and the spectrophotometry is accurate to 10\%.

The objects from reference 7 are the brightest galaxies from the 
Calan-Tololo objective prism survey \citep{maza89}. Its limitinig magnitude is 17.5.
The spectra were obtained using a variety of detectors (Vidicon, 2DF, CCDs), and the apertures used
range from 2\arcsec to 4\arcsec. The spectra were flux calibrated.

The HII galaxies from reference 19 are located in the voids of the digitized 
Hamburg Quasar Survey \citep{hagen95}. The limiting magnitude of the objective prism survey
is 18.5. The data presented in this work were gathered using the 2.2m telescope at
the German-Spanish observatory at Calar Alto, Spain, under good photometric conditions
using a 4\arcsec slit. This is enough to encompass most of the line-emitting region.

The objects from the reference 15 were selected from the Case survey \citep{pesch83}.
This survey searches simultaneously for both a UV excess and strong emission lines. The 
limiting magnitude is 16.0. The data for this spectroscopic follow-up study were obtained on 9 observing runs, using
different telescopes and detectors. The majority of the objects in this sample were observed 
using CCD detectors, and the spectra were flux calibrated. The slit widths used were 2.4\arcsec 
and 3.0\arcsec. The spatial region extracted spanned all of the line emission, but did not cover all 
of the continuum emission.

The objects from the references 13, 14, 16, 17 and 18  were taken from the first
Byurakan Survey (FBS, also known as the Markarian survey,\citep{mrk67}) and the second Byurakan 
Survey (SBS)\citep{sbs}. The FBS is another objective prism survey that searches for galaxies with a 
UV excess, and its limiting photographic magnitude is 15.5.
Selection of the SBS objects was done according to the presence of strong UV continuum and emission 
lines. The SBS was carried out using the same Schmidt telescope as the FBS, and its limiting 
photographic magnitude is 19.5.
The observations presented in 13, 14, 16 and 17 are high S/N spectra, taken with the Ritchey-Chretien
 spectrograph at the Kitt Peak National Observatory (KPNO) 4m telescope, with the T2KB CCD. The slit 
width used was 2\arcsec, and the nights were transparent.
The spectroscopic observations from reference 18 were done using the Ritchey-Chretien spectrograph at the
KPNO 4m telescope, and with the GoldCam spectrograph at the 2.1 m KPNO telescope. In the majority of 
cases, a 2\arcsec slit was used.

Initially, we compiled objects with data on, at least, the intensities of the following emission lines:
[OII] $\lambda\lambda$ 3727,29 \AA\ , [OIII] $\lambda$ 4363 \AA\ , [OIII] $\lambda\lambda$ 4959, 5007 \AA\ ,
and [SII] $\lambda\lambda$ 6717, 6731 \AA\ . The hydrogen Balmer recombination lines were also
required in order to allow a proper reddening correction. The emission lines of [SII] allow the 
determination of the electron
density (see e.g. \citealt{ost89}) and can also be used, as well as the lines of [OII], to estimate the
ionization parameter of the emitting gas. The intensities of the auroral and nebular 
[OIII] emission lines are needed in order to derive accurate values of the electron temperature,
and hence of the oxygen abundance. Objects meeting these requirements belong mostly to references 6, 
10, 13, 14 and 16. For all the objects in these samples, data on the
intensity and equivalent width of H$\beta$ are provided. We have also
included data from references 7 and 2, although they lack data on the H$\beta$ 
equivalent width and line intensity respectively.

Our initial sample was later enlarged to include objects with the same information as above but with no data on the
[SII] lines, for which a low density regime was assumed. This is probably not a bad assumption since the average
electron density derived for the galaxies with [SII] data is around 200
\textrm{cm}$^{-3}$.

Finally, a third enlargement was made to include objects with no
reported measurements of the [OIII] $\lambda$ 4363 \AA\
line. These may correspond to objects with low surface brightness and/or low excitation and are generally
excluded from emission-line analyses of samples of HII galaxies since the derivation of their oxygen abundances require the use
of empirical calibrations which are rather uncertain (see e.g. \citealt{Ski89}). These 
objects represent, however,
a large fraction of the observed HII galaxies. Most of them have been
taken from references 6, 19 and 15.
The latter reference does not provide absolute intensities for the hydrogen recombination lines.

The final sample comprises 450 objects: 236 with data on the [OIII] $\lambda$ 4363 \AA\
line and 214 with no observations of this line. These data have been obtained according to different selection criteria and
using different instruments and techniques, and the parent populations of the different samples used are different. 
The presented sample can't be considered as complete in any sense. In particular, line selected samples
of galaxies are complete to a given line+continuum flux, whereas color selected samples will be complete to a given
apparent magnitude. The galaxy sample presented here constitutes, however, the largest sample of local 
HII galaxies with good quality spectroscopic data, to our knowledge. However, the sample is very inhomogenous, due 
to different instrumental setups, observing 
conditions and reduction procedures. An accuracy analysis is therefore needed. This was done by comparing the 
observations for a few very well studied objects -- {\it e.g.} IZw18, IIZw40, Mk36 -- for which several 
independent observations exist. We have treated them as individual data in order to examine the external
observational errors. The average error in redshift determinations is 5\%. The average error in H$\beta$ fluxes
is 45\%, and the average error in the equivalent width of H$\beta$, W$_{\beta}$ is 16\%.
Bin widths in the forthcoming histograms were chosen to be wide enough to engulf these errors.
It is also important to note that, however large these numbers seem to be, they were derived from the 
nearest, brightest and best studied objects. These sources are sensitive to the full range of the aforementioned 
uncertainities, and are particularly affected by aperture issues (several components, different position angles for the slit, etc).
More distant objects will be less affected by such effects, and their measurements will probably be more 
accurate. The errors previously presented are likely to be upper limits to the real uncertainities.

Table \ref{m1y2pres} lists the emission line properties of the
sample objects. Column 1 gives the name of the object as it appears in
the reference indicated in column 2. Column 3 gives the galaxy redshift
({\it cz}). Columns 4 to 9 give the reddening corrected emission line
intensities, relative to that of H$\beta$, of: [OII] $\lambda\lambda$ 3727,29 \AA,
[OIII] $\lambda$ 4363 \AA, [OIII] $\lambda$ 4959 \AA, [OIII] $\lambda$ 5007 \AA, [SII] $\lambda$
6717 \AA{} and [SII] $\lambda$ 6731 \AA. Column 10 gives the value of the
logarithmic extinction at H$\beta$, c(H$\beta$). Column 11 gives the
reddening corrected H$\beta$ flux, in \textrm{ergs cm}$^{-2}$ \textrm{s}$^{-1}$. Finally,
columns 12 and 13 give the equivalent width of the H$\beta$ and [OIII] $\lambda$
5007 \AA{} lines in \AA. Only an example of the table entries are shown, the
complete table being available in the online version of the paper.

\onecolumn
\begin{landscape}


\begin{table*}
\begin{minipage}{175mm}
\caption{Partial list of galaxies included in this study.} 
\label{m1y2pres}

\begin{tabular}[l]{|ll|llllllllllll|}

Object. &  Reference. & \textit{cz} & $\lambda3727$  & $\lambda4363$ & $\lambda4959$ & $\lambda5007$ &
$\lambda6584$  &  $\lambda6716$ & $\lambda6731$ & c(H$\beta$) & 
$\log \mathrm{F(H}\beta\mathrm{).}$ & $W_{\beta}$ & $W_{\lambda5007}$ \\

\hline
UM439	&	6	&	1199	&	0.846	&	0.125	&	2.611	&	7.732	&	0.037	&	0.093	&	0.061	&	0.00	&	-13.44	&	160	&	1325  \\
UM448	&	6	&	5396	&	2.736	&	0.034	&	0.961	&	2.826	&	0.392	&	0.277	&	0.197	&	0.27	&	-12.98	&	43	&	127  \\
UM448	&	17	&	5498	&	2.776	&	0.030	&	0.867	&	2.599	&	0.409	&	0.366	&	0.285	&	0.33	&	-12.60	&	49	&	{\ldots}  \\
UM461	&	6	&	899	&	0.455	&	0.157	&	2.136	&	6.434	&	0.020	&	0.040	&	0.032	&	0.00	&	-13.20	&	342	&	2254  \\
UM461	&	17	&	1007	&	0.527	&	0.136	&	2.039	&	6.022	&	0.021	&	0.052	&	0.042	&	0.12	&	-13.47	&	223	&	{\ldots}  \\
UM463	&	10	&	1199	&	1.255	&	0.153	&	1.942	&	5.687	&	0.078	&	0.088	&	0.091	&	0.05	&	-13.77	&	127	&	716  \\
UM462SW	&	6	&	899	&	1.599	&	0.101	&	1.896	&	5.660	&	0.060	&	0.097	&	0.081	&	0.09	&	-13.24	&	124	&	781  \\
UM462SW	&	17	&	1028	&	1.742	&	0.078	&	1.663	&	4.929	&	0.073	&	0.168	&	0.112	&	0.29	&	-13.02	&	100	&	{\ldots}  \\
UM462knotA	&	10	&	899	&	1.488	&	0.102	&	2.023	&	6.044	&	0.071	&	0.114	&	0.086	&	0.03	&	-13.38	&	149	&	{\ldots}  \\
Tol1156-346	&	6	&	2398	&	0.913	&	0.123	&	2.717	&	7.924	&	0.054	&	0.068	&	0.057	&	0.58	&	-13.64	&	111	&	978  \\
UM483	&	10	&	2398	&	2.442	&	0.053	&	1.982	&	5.925	&	0.080	&	0.121	&	0.049	&	0.50	&	-13.85	&	27	&	181  \\
Tol1304-353	&	10	&	4197	&	0.416	&	0.195	&	2.240	&	6.837	&	{\ldots}	&	0.034	&	0.014	&	0.00	&	-13.41	&	253	&	1929  \\
Tol1324-276	&	6	&	1798	&	1.456	&	0.050	&	1.915	&	5.568	&	0.100	&	0.123	&	0.105	&	0.23	&	-13.11	&	113	&	652  \\
Tol1327-380	&	6	&	7794	&	3.233	&	0.037	&	2.067	&	5.889	&	0.103	&	0.128	&	0.085	&	1.14	&	-13.85	&	53	&	362  \\
NGC5253a	&	6	&	300	&	1.296	&	0.066	&	1.597	&	4.800	&	0.223	&	0.127	&	0.105	&	0.00	&	-12.15	&	216	&	0.00  \\
Tol1345-420	&	6	&	2398	&	1.765	&	0.073	&	1.853	&	5.476	&	0.051	&	0.145	&	0.137	&	0.17	&	-13.85	&	67	&	391  \\
Tol1400-411	&	6	&	600	&	0.957	&	0.121	&	2.296	&	6.856	&	0.043	&	0.072	&	0.055	&	0.05	&	-12.85	&	259	&	1899  \\
SBS1420+544*	&	17	&	6176	&	0.577	&	0.184	&	2.263	&	6.862	&	0.019	&	0.057	&	0.028	&	0.16	&	-13.55	&	217	&	{\ldots}  \\
SBS1533+469	&	14	&	5666	&	2.249	&	0.077	&	1.561	&	4.854	&	0.154	&	0.300	&	0.220	&	0.04	&	-13.85	&	30	&	{\ldots}  \\
\hline
Mrk507	&	6	&	5996	&	4.261	&	{\ldots}	&	1.896	&	4.884	&	{\ldots}	&	{\ldots}	&	{\ldots}	&	1.04	&	-14.03	&	54	&	302 \\
Tol2122-408	&	6	&	4197	&	4.901	&	{\ldots}	&	1.577	&	3.931	&	{\ldots}	&	{\ldots}	&	{\ldots}	&	0.82	&	-13.89	&	13	&	56 \\
UM162	&	6	&	19786	&	2.171	&	{\ldots}	&	2.358	&	7.018	&	{\ldots}	&	{\ldots}	&	{\ldots}	&	0.06	&	-14.19	&	81	&	600 \\
UM3	&	6	&	6296	&	2.425	&	{\ldots}	&	1.107	&	3.383	&	{\ldots}	&	{\ldots}	&	{\ldots}	&	0.44	&	-14.03	&	33	&	120 \\
UM4W	&	6	&	6296	&	2.129	&	{\ldots}	&	1.535	&	4.551	&	{\ldots}	&	{\ldots}	&	{\ldots}	&	0.04	&	-13.92	&	30	&	144 \\
UM9	&	6	&	5096	&	2.816	&	{\ldots}	&	1.672	&	4.442	&	{\ldots}	&	{\ldots}	&	{\ldots}	&	{\ldots}	&	-14.39	&	15	&	74 \\
UM191	&	6	&	7195	&	4.790	&	{\ldots}	&	0.337	&	0.727	&	0.826	&	{\ldots}	&	{\ldots}	&	1.33	&	-13.85	&	10	&	8 \\
Mrk109	&	2	&	9100	&	3.810	&	{\ldots}	&	0.426	&	1.370	&	0.763	&	0.519	&	0.271	&	0.36	&	-13.90	&	27	&	{\ldots} \\
Mrk168	&	2	&	10148	&	3.700	&	{\ldots}	&	1.440	&	4.370	&	0.281	&	0.301	&	0.233	&	0.63	&	-13.95	&	39	&	{\ldots} \\
NGC3690	&	2	&	3121	&	2.600	&	{\ldots}	&	0.427	&	1.270	&	0.968	&	0.288	&	0.216	&	0.73	&	-12.26	&	32	&	{\ldots} \\
M03.13	&	7	&	10490	&	2.570	&	{\ldots}	&	1.259	&	3.548	&	0.240	&	0.269	&	0.295	&	0.06	&	-14.11	&	{\ldots}	&	{\ldots} \\
CG34	&	15	&	5126	&	2.094	&	{\ldots}	&	1.610	&	4.879	&	{\ldots}	&	{\ldots}	&	{\ldots}	&	0.18	&	{\ldots}	&	52	&	275 \\
CG74	&	15	&	13910	&	4.065	&	{\ldots}	&	1.332	&	4.037	&	{\ldots}	&	{\ldots}	&	{\ldots}	&	0.88	&	{\ldots}	&	21	&	103 \\
CG85	&	15	&	14900	&	4.722	&	{\ldots}	&	1.141	&	3.458	&	{\ldots}	&	{\ldots}	&	{\ldots}	&	0.82	&	{\ldots}	&	20	&	79 \\
CG103	&	15	&	1619	&	3.800	&	{\ldots}	&	0.967	&	2.929	&	{\ldots}	&	{\ldots}	&	{\ldots}	&	0.54	&	{\ldots}	&	31	&	95 \\
CG136	&	15	&	6895	&	3.529	&	{\ldots}	&	1.358	&	4.116	&	{\ldots}	&	{\ldots}	&	{\ldots}	&	0.19	&	{\ldots}	&	30	&	121 \\
CG141	&	15	&	64456	&	1.910	&	{\ldots}	&	1.544	&	4.680	&	0.143	&	{\ldots}	&	{\ldots}	&	0.00	&	{\ldots}	&	56	&	269 \\
CG147	&	15	&	3328	&	5.026	&	{\ldots}	&	0.981	&	2.973	&	0.429	&	{\ldots}	&	{\ldots}	&	0.93	&	{\ldots}	&	21	&	62 \\
\hline


\end{tabular}
\medskip

References to the table: 1,\citet{lequ79}; 2,\citet{kunth83}; 3,\citet{french80}; 4,\citet{dinner86}; 5,\citet[]{moles90}; 
6,\citet{rjt91}; 7,\citet{pena91}; 8,\citet{pagel92.orig}; 9,\citet{ski93}; 10,\citet{masegosa94}; 11,\citet{ski94}; 
12,\citet{sarsearl72}; 13,\citet{izo94}; 14,\citet{thuan95}; 15,\citet{salz95}; 16,\citet{izo97}; 17,\citet{izo98}; 
18,\citet{guseva00} and 19, \citet{popes00}.


\end{minipage}
\end{table*}

\end{landscape}


\twocolumn

\section{Statistical properties of the general sample}

In the study of the ionised medium of star forming regions, the detection of
the weak auroral line [O\textsc{III}]$\lambda 4363$ \AA{} constitutes an
important source of information. It allows the derivation of  accurate electron temperatures and hence
oxygen abundaces. Therefore, it is not surprising that most works on
the ionized nebulae of H\textsc{II} galaxies focus on objects with data on this valuable
line. In many sources however, this line is not present. In nearby
objects, its proximity to the Balmer H$\gamma$ line which often shows prominent absortion wings, and the
sky Hg $\lambda$ 4359 \AA{} line complicates its measurement and the line is
intrinsically weak in objects with a relatively low excitation. On the
other hand, in the distant universe, very few sources are expected to
show the line since the cosmological dimming factor
$(1+z)^{4}$, which at a redshift of only 0.4 is about 4, makes its
detection very difficult. The study of the properties of sources not showing the [O\textsc{III}]$\lambda
4363$ \AA{} line is therefore of great importance in order to provide an
adequate comparison sample to study the distant population of
blue compact objects.

For our study we have split the total HII galaxy sample into two
subsamples. Subsample Sub1 comprises 236 
objects with measurements of the [OIII]$\lambda 4363$ \AA{} line
intensity. Subsample Sub2 comprises objects for which
the [OIII]$\lambda 4363$ \AA{} line is not reported or is too weak to be measured. This
latter subsample consists of 214 objects.

\subsection{Subsample characterization.}

In order to get a picture of the observable properties of both subsamples, the distributions of the
following quantities were drawn.

\begin{itemize}
\item[(a)]{Observed H$\beta$ flux, F(H$\beta$).}
\item[(b)]{Radial velocity, $cz$.}
\item[(c)]{Extinction, c(H$\beta$).}
\item[(d)]{H$\beta$ luminosity, L(H$\beta$).}
\item[(e)]{H$\beta$ equivalent width, W$_{\beta}$.}
\end{itemize}

\begin{figure}
\includegraphics[scale=0.5]{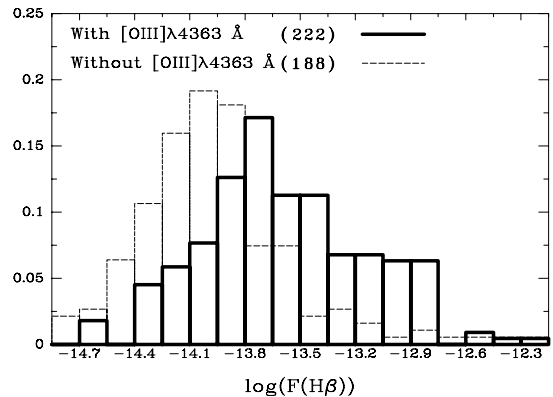}
\caption{\small \sl \emph{Observed} H$\beta$ flux distributions.}
\label{fhb_d}
\end{figure}

\begin{figure}
\includegraphics[scale=0.5]{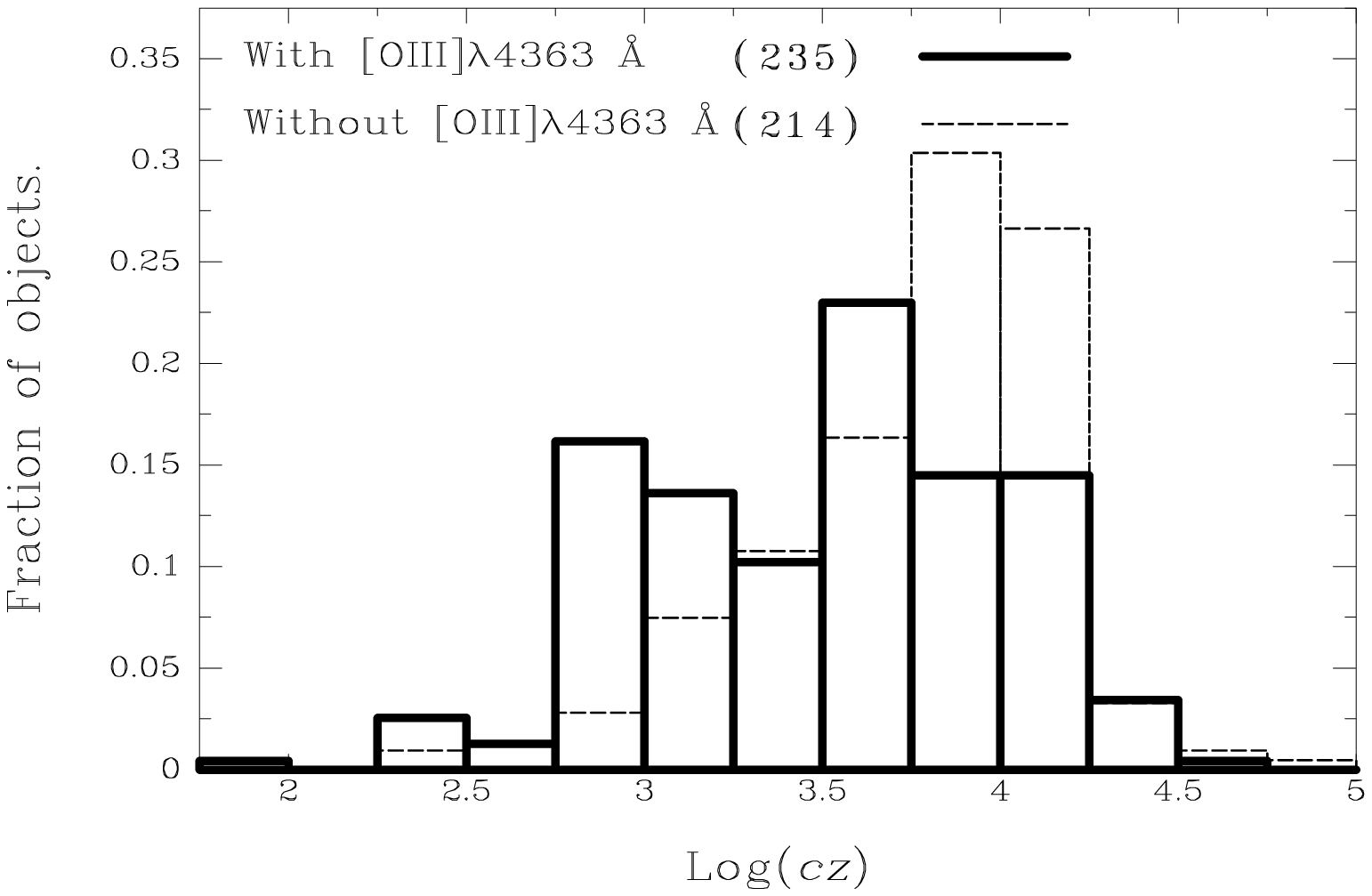}
\caption{\small \sl Radial velocity distributions.}
\label{cz_d}
\end{figure}

\begin{figure}
\includegraphics[scale=0.5]{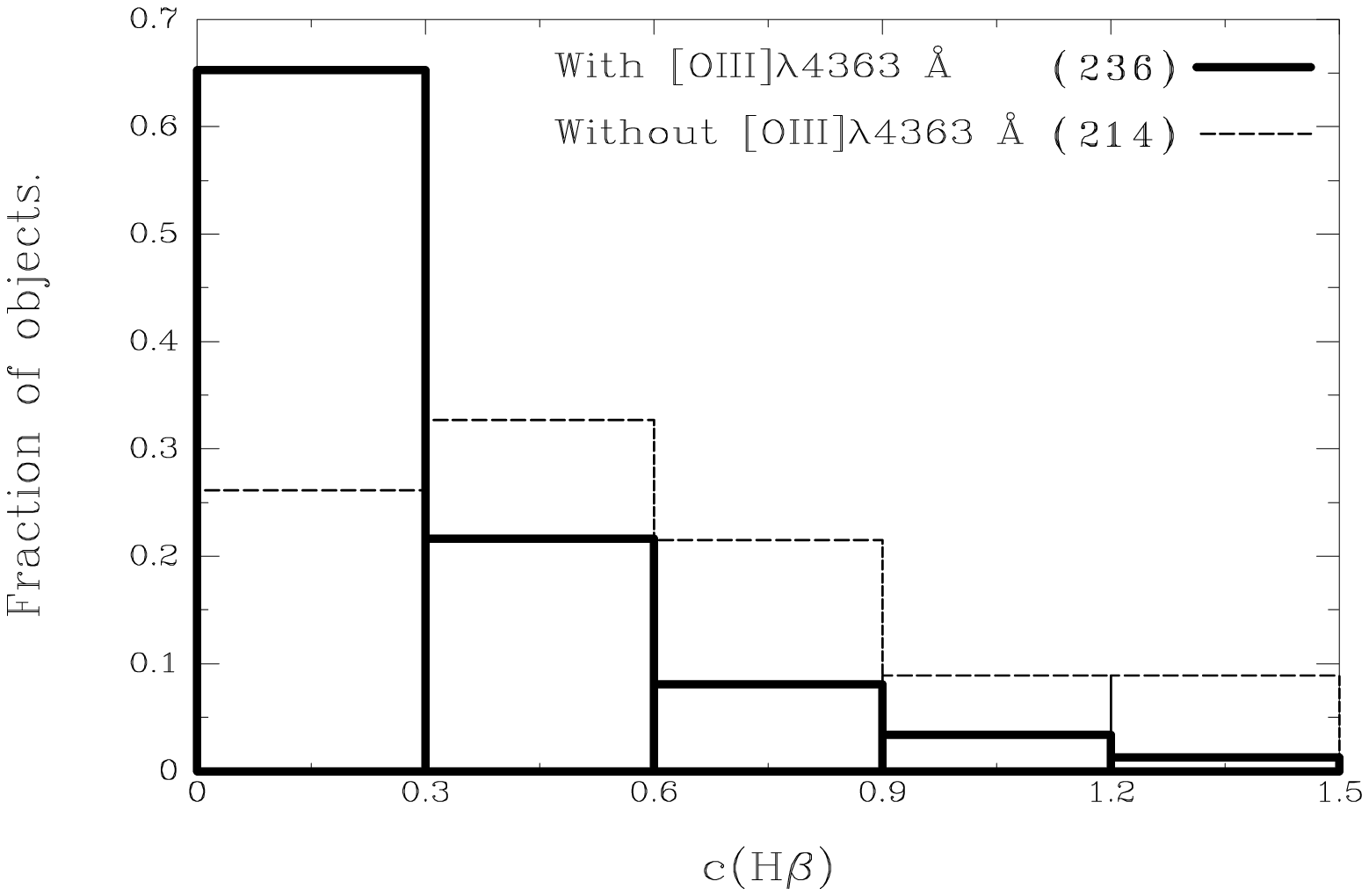}
\caption{\small \sl c(H$\beta$) distributions.}
\label{chb_d}
\end{figure}

\begin{figure}
\includegraphics[scale=0.5]{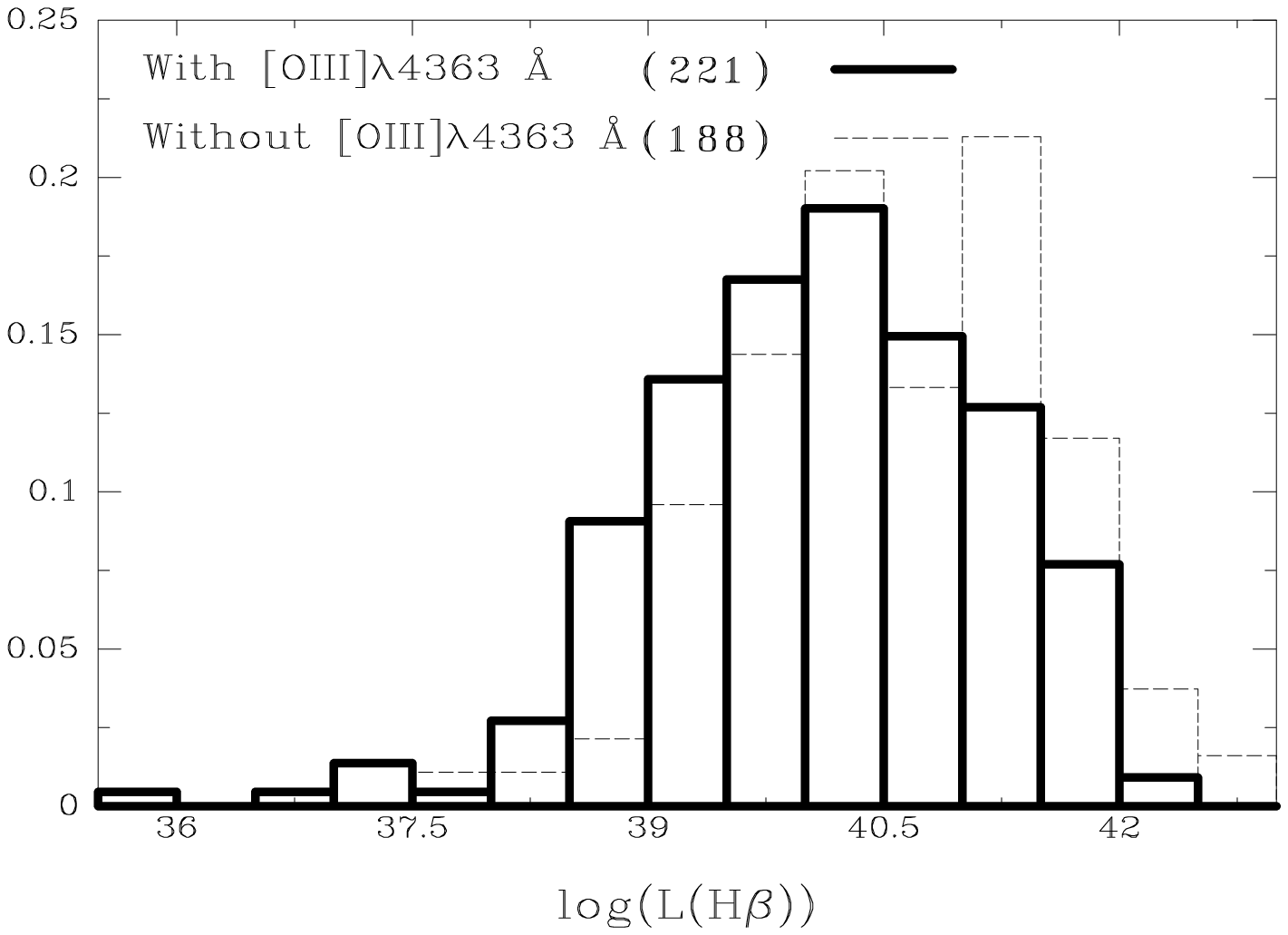}
\caption{\small \sl H$\beta$ luminosity distributions.}
\label{lglhb_d}
\end{figure}

\begin{figure}
\includegraphics[scale=0.5]{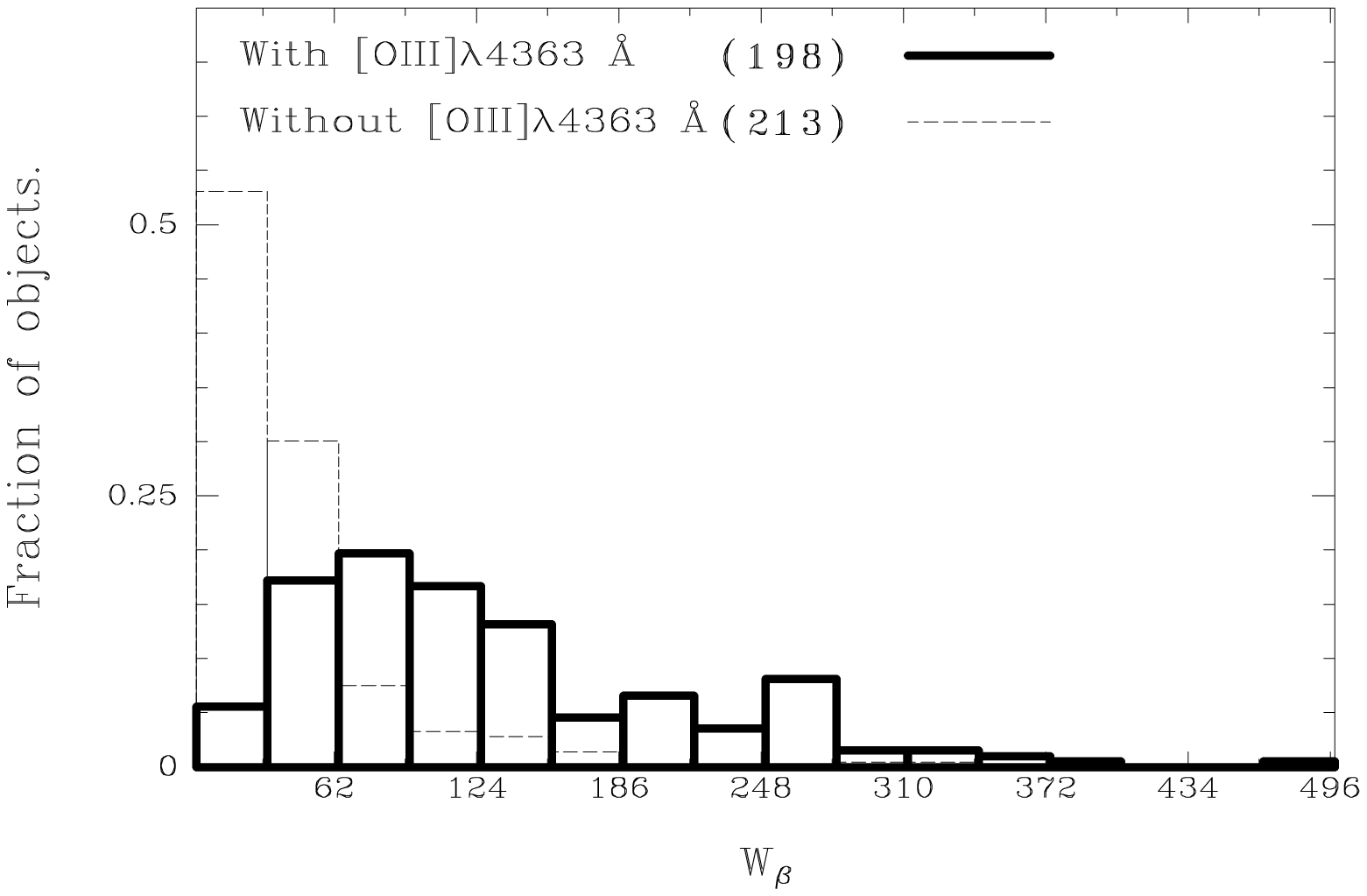}
\caption{\small \sl W$_{\beta}$ distributions.}
\label{wb_d}
\end{figure}

The corresponding histograms are presented in figures \ref{fhb_d} through
\ref{wb_d}. The H$\beta$ flux and W$_{\beta}$ distributions
were drawn using directly the reported
data from the literature. The radial velocities were derived from the
reported \textit{z} values. However, in the cases in which this number was 
not given in the literature, the distance
value from the NED database was converted to radial velocities using the cosmology 
given above. The H$\beta$ luminosities were derived from the
luminosity-distances and from the extinction-corrected H$\beta$ fluxes.
We have neglected the solar system velocity with respect to the CMB, which is 370 
\textrm{km s}$^{-1}$. This effect can only affect the luminosity calculations
of the nearest objects. However, for the vast majority of sources, this
does not introduce a big error since their radial velocities are much greater.
Furthermore, this additional error is engulfed by the $\log L(H\beta)$
histogram bin width.
In all cases, the reddening was re-estimated from the available Balmer
decrements. For the objects from reference 6 affected by second order contamination, only
the objects for which the H$\gamma$/H$\beta$ decrement was available were selected. In some 
cases however, the nature of the data did not allow an accurate determination. Therefore, in order to
minimize any spurious effects, objects with estimated values of the
logarithmic extinction at H$\beta$, c(H$\beta$), larger than 1.5 were
excluded from our analysis. Also, it is found that the extinction is 
low for most objects. For this reason, even if the determination of $c(H\beta)$ is 
affected by several issues such as underlying absorptions or the reddening 
curve adopted, these factors will not have a great impact on derived quantities such as 
line luminosities or the [OIII]$\lambda \lambda 4959+5007$/[OII]$\lambda 3727$ ratio.
Simple statistics on the presented distributions are given in table \ref{tabla}.
Table \ref{tabla} also gathers the estimated error in each variable, and
the bin width of the corresponding histogram. It is seen that the bin width is, in all cases, at
least equal to the given error. We think that the conclusions drawn from these histograms are robust.


{\tiny
\begin{flushleft}
\begin{table*}
\caption{\small \sl General properties of the galaxies studied.} 
\begin{tabular}[t]{|ccc|ccc|ccc|} \hline
                    & &  &  With   &        &        & Without &        \\ \hline
                    & Error & Bin Width   &  Mean   & Median & Standard deviation & Mean  & Median  & Standard deviation  \\ \hline

$\log L(H\beta)$   & 0.30 & 0.50 &  40.1   &  40.1   & 1.1    & 40.6  & 40.6    & 1.0    \\ 
c(H$\beta$)         & 0.30 & 0.30 &  0.28   & 0.20    & 0.29   & 0.58  & 0.50    & 0.43   \\
$\log(cz)$          & 0.02 & 0.25 &  3.50   &  3.56   & 0.51   & 3.78  & 3.84    & 0.39   \\
$W_{\beta}$         & 16   & 32   &  131    & 110    & 86     & 46    & 31      & 51.6   \\
$\log F(H\beta)$    & 0.15 & 0.15 &  -13.6  &  -13.6  & 0.45   & -14.0 & -14.0   & 0.43   \\


\end{tabular}
\label{tabla}
\end{table*}
\end{flushleft}
}

Several points can be made from the presented histograms.
Figure \ref{fhb_d} shows that the observed H$\beta$ flux from galaxies without measurements of the 
[O\textsc{III}]$\lambda$ 4363 \AA\ is
lower than the observed flux for galaxies with data on this line. 
There is an evident excess of Sub2 members at low observed fluxes.
On average, the flux from Sub1 objects is 2.4 times greater than 
that of Sub2 members. In a number of cases, the undetection of the
[O\textsc{III}]$\lambda 4363$ line will not be due to its weakness relative
to other lines, but due to the general faintness of the observed 
emission line spectrum.
Figures \ref{cz_d} and \ref{chb_d} show the redshift and extinction distributions for
both subsamples. It is seen that objects presenting the [O\textsc{III}]$\lambda 4363$ 
line are located at lower distances than Sub2 members. This will make the
[O\textsc{III}]$\lambda$ 4363 line easier to detect. Also, sources with the 
[O\textsc{III}]$\lambda 4363$ line seem to be less affected 
by extinction. The majority of objects presenting the auroral line have
logarithmic extinctions lower than 0.5, while that is the average value of
the extinction coefficient for subsample 2 sources. In some cases, the 
auroral line will be absorbed by dust and rendered unobservable.
Figure \ref{chb_d} also shows that most H\textsc{II} galaxies present low 
extinctions ($c(H\beta)\leq 0.6$), implying that extinction will not be an important 
ingredient in the error budget.
Finally, figure \ref{lglhb_d} shows that, in spite of all these differences the 
H$\beta$ luminosity distributions are very similar for both 
subsamples, Sub2 objects being marginally more luminous.
Regarding ionisation properties, figure \ref{wb_d} shows that  
Sub1 objects show higher H$\beta$ equivalent widths. The H$\beta$
equivalent width distributions present a different shape too. This points to
an evolutionary effect, in the sense that the ionising clusters of the objects not showing 
the [O\textsc{III}]$\lambda 4363$ line might be more evolved.

\subsection{The mass of the ionising cluster.}

The mass of the ionizing cluster was calculated from the H$\beta$ luminosities and
equivalent widths using the following
considerations.\par
As the star cluster ages, the number of hydrogen ionizing photons per unit
mass of the ionizing cluster decreases. Assuming that the age of the
ionizing cluster is related to W$_{\beta}$, a relation should exist
between the number of hydrogen ionizing photons per unit
mass of the star cluster and W$_{\beta}$. Such relation is given for
single-burst models in \citet{diaz99}. The expression used is
\[\log(Q(H)/M_{*})=44.8+0.86\times\log W_{\beta,0}\]
In this relation,
W$_{\beta,0}$ is the equivalent width, in angstroms, that would be observed in the absence of
an underlying population, Q(H) is the number of hydrogen ionizing photons per second and M$_{*}$
is expressed in solar masses. \par
\begin{figure}
\includegraphics[scale=0.5]{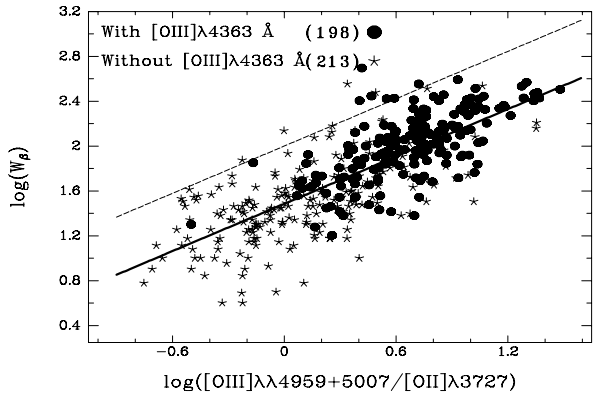}
\caption{\small \sl W$_{\beta}$ \textit{vs.} excitation relation.}
\label{wb.o3_o2_r}
\end{figure}

\begin{figure}
\includegraphics[scale=0.5]{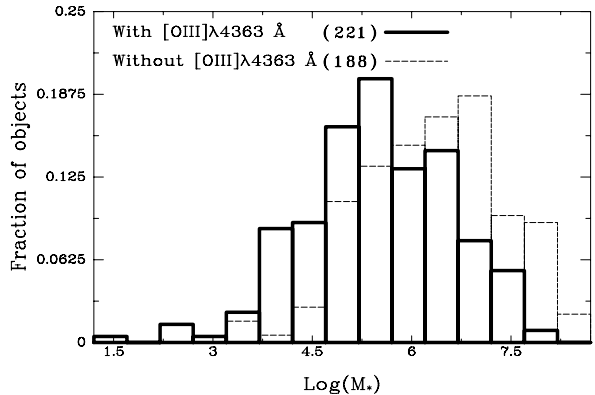}
\caption{\small \sl Cluster mass distributions.}
\label{mc_d}
\end{figure}

 No direct information about the presence or absence of an underlying stellar 
populations exists. However, a well defined relation exists
between  W$_{\beta}$ and the degree of ionization of the nebula, albeit showing a 
scatter larger than observational errors.
In figure \ref{wb.o3_o2_r}, W$_{\beta}$ is plotted as a function of 
the [OIII]$\lambda \lambda 4959+5007$/[OII]$\lambda 3727$ ratio.
If the degree of ionization is ascribed 
to the age of the ionizing star cluster, it is reasonable to assume that 
the vertical scatter shown by the data is due to different contributions of continuum light from 
underlying populations. A least squares fit is presented as a solid 
line. The dashed line represents an upper
envelope to the data which is located 1.5 times the \textit{rms} above the fit 
and corresponds to the relation:
\[ \log W_{\beta ,0}=2.000+0.703\log \frac{\mathrm{[OIII]}}{\mathrm{[OII]}}  \]
The objects located on this upper envelope are likely to be the ones in which
the underlying population is minimum. We have used 
this upper envelope to calculate the mass of the ionizing cluster.
There is, however, a metallicity effect in figure \ref{wb.o3_o2_r}. In the
absence of any underlying stellar population, H\textsc{II} galaxies with higher 
metalicity clouds will present a lower 
[OIII]$\lambda \lambda 4959+5007$/[OII]$\lambda 3727$ ratio due to lower
effective temperatures of their ionizing stars (\citet{epm_diaz_arbitro} and \citet{kew02}).
However, there will be no change in the equivalent width of H$\beta$, to 
zero-th order. Therefore, $W_{\beta,0}$ should be a 
function of both the ionization ratio and metallicity. However, only the 
[OIII]$\lambda \lambda 4959+5007$/[OII]$\lambda 3727$ will be used here.
At any rate, the derived cluster mass constitutes only a lower limit 
since some photons might be actually escaping from the nebula, or there might 
be dust globules within the ionized medium. If a significant fraction of the ionizing photons
escape from the nebula unprocessed, the ionizing star cluster will appear to be less massive 
than it really is. On the other hand, if the dust optical depth is large within the 
nebula, the absorbed energy will be re-emitted at other wavelengths, and the star cluster mass will be
underestimated again. Observations at other wavelengths would be required to quantify these 
issues.  
Figure \ref{mc_d} presents the histograms of the ionizing cluster masses for both 
subsamples. The estimated error in $\log M_{*}$ is 0.40, and the bin width is 0.5.


\subsection{The [OIII]$\lambda \lambda 4959+5007$/[OII]$\lambda 3727$ distributions.}

\begin{figure}
\includegraphics[scale=0.5]{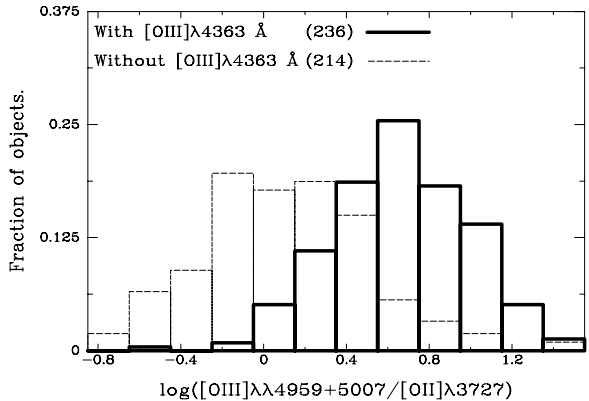}
\caption{\small \sl [OIII]$\lambda \lambda 4959+5007$/[OII]$\lambda 3727$ distributions.}
\label{o3_o2_d}
\end{figure}

Figure \ref{o3_o2_d} shows the [OIII]$\lambda \lambda 4959+5007$/[OII]$\lambda 3727$ distributions.
The estimated error in the ionization ratio is 0.2dex, and the bin is 0.2dex wide. It is seen that
galaxies without the [OIII]$\lambda 4363$ line show lower ionization ratios. 
This separation in [OIII]$\lambda \lambda 4959+5007$/[OII]$\lambda 3727$ can't be explained
by uncertainties in the reddening determination, since the typical differences in
c$(H\beta)$ observed between both subsamples would only change the ionization
ratio by 0.1dex, much less than the observed separation of 0.6dex.
The ionization ratio was correlated 
with W$_{\beta}$ in figure \ref{wb.o3_o2_r}, which is now re-examined in order to 
gain insight on what effects could be responsible 
for the segregation observed in figure \ref{o3_o2_d}. We begin commenting on 
two possible evolutionary scenarios:
\begin{enumerate}
\item{Long-term evolution. In the framework of succesive bursts scenario, as
the continuum level and the amount of coolants rise, the O$^{++}$ auroral line
becomes fainter and might be swallowed under the continuum noise or stellar
features. Objects without the auroral line would therefore show lower equivalent 
widths. Less luminous \textit{starbursts} are not required in this scenario.}
\item{Short-term evolution of the \textit{starburst}. The W$_{\beta}$ of
older clusters is lower than that of younger ones simply because they are
less able to produce ionizing photons. This would make the [OIII] line naturally weak and
unobservable even in the presence of a not-too-strong underlying population.
This is likely to play a major role. Additionally, high-metallicity objects
and/or very evolved ones might have a lower effective temperature which
would produce a lower [OIII]$\lambda\lambda4959+5007$/[OII]$\lambda 3727$ 
at comparable ionization parameter. Figure 
\ref{wb.o3_o2_r} shows the W$_{\beta}$ \textit{vs.} 
[OIII]$\lambda \lambda 4959+5007$/[OII]$\lambda 3727$ diagram. It can be seen 
that the observed galaxies, with and without [OIII]$\lambda$ 4363 are 
separated. The two subsamples lie along the relationship, but on opposite ends of it. This
suggests that Sub1 objects are tipically younger than Sub2 subsample objects,
according to this picture.}
\end{enumerate}
The real picture, of course, will be a combination of the two effects.

\begin{figure}
\includegraphics[scale=0.5]{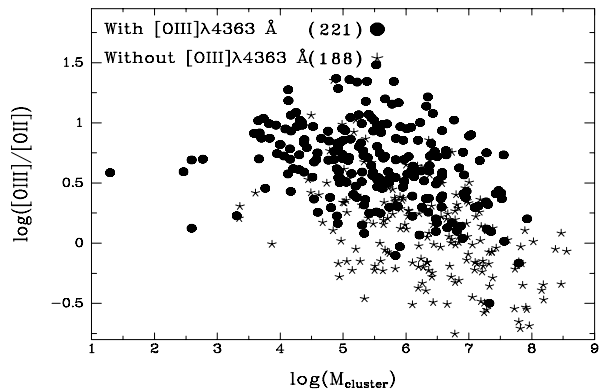}
\caption{\small \sl Ionization ratio \textit{vs.} Cluster mass relation.}
\label{logo3_o2.vs.logMc}
\end{figure}

Other, non-evolutionary possible reasons as to why the ionization ratio
[OIII]$\lambda\lambda4959+5007$/[OII]$\lambda 3727$ may be lowered are:
\begin{itemize}
\item{The ionization parameter depends on the mass of the ionizing 
cluster, since more massive clusters will harbour a greater number of massive stars.
However, it can be seen in figure \ref{mc_d} that the distribution of 
cluster masses for both subsamples greatly overlap. Furthermore, figure
\ref{logo3_o2.vs.logMc} shows that objects without [O\textsc{III}]$\lambda$4363
have a lower ionization ratios than objects with the auroral line even
for the most massive clusters. This indicates that cluster mass is not
responsible for this segregation in
[OIII]$\lambda\lambda4959+5007$/[OII]$\lambda 3727$.\footnote{And it also
indicates that subsampling of the Initial Mass Function, which lowers
the ionization ratio (the so called \textit{richness effect}) may not account
for all the separation.}}
\item{If the electron density is higher than the [OII]$\lambda$3727 transition
critical density, the ionization ratio is reduced. In particular, its value at
$N_{e}=10^{4}$\textrm{cm}$^{-3}$ is 3.75 times smaller than its value at
$N_{e}=200$\textrm{cm}$^{-3}$.
Since H\textsc{II} galaxies have an electron density which is well below
the critical density, this option is not likely to be responsible for the 
observed differences in [OIII]$\lambda\lambda4959+5007$/[OII]$\lambda 3727$.
However, it is a possibility for some objects.}
\item{Geometry, dust content and photon escape. If an important fraction of the ionizing
photons escape the nebula, the ionization parameter is low because
of geometrical reasons, or a significant fraction of high-energy photons
are absorbed by dust within the nebula, this would weaken the $O^{++}$ auroral line.
These phenomena might be partially responsible for the segregation observed in  the 
[OIII]$\lambda\lambda4959+5007$/[OII]$\lambda 3727$ histogram.}
\end{itemize}

\section{Metallicity analysis.}

The detection or undetection of the [O\textsc{III}]$\lambda4363$ 
line is affected by several purely observational factors such as distance to the observed 
galaxy, galactic extinction, quality of the spectrum, etc.\ that should not correlate 
with the oxygen abundance. However, high oxygen abundances, implying low electron 
temperatures in the gas, could render the line too weak to be detected since
the intensity of this auroral line depends on electron temperature. There are then 
two extreme cases.

\begin{enumerate}

\item{The objects in  subsample Sub2 are, on average,  more metal-rich than the Sub1
objects. The [O\textsc{III}] auroral line would then be naturally 
weak, and the absence of this line in a particular spectrum would bias the oxygen 
abundance of the ionized gas of a given galaxy towards higher metallicities. There 
should be a separation in excitation  between subsamples Sub1 and
Sub2 in this case. It is seen in figure \ref{o3_o2_d} that this is indeed 
the case.However, it has to be borne in mind that a simple age effect would produce the same
effect on the excitation and equivalent width distributions.}

\item{The objects which do not show the [OIII]$\lambda 4363$ line have higher
extinctions, their continuum is stronger or are affected by other observational 
problems. This may cause the auroral line to be
swallowed in the continuum noise. The undetection of the auroral line would then
be an observational issue only, and the presence or absence of this line in the spectrum 
of a galaxy should not bias its metallicity in any way.}
\end{enumerate}

Since our present knowledge of the metallicity distribution of H\textsc{II} galaxies
comes from the analysis of samples for which the [O\textsc{III}]$\lambda4363$ line is 
observed, it is important to check if the metallicity distributions of subsamples 
Sub1 and Sub2 really differ. 

\subsection{Metallicity analysis for Sub1 objects.}
For subsample Sub1, the oxygen abundance was derived through the 
determination of the electron temperature using the
[O\textsc{III}]$\lambda$4363/[O\textsc{III}]$\lambda$5007 line ratio.
The procedure can be reviewed in \citet{pagel92.orig}.
The density was estimated via the 
[S\textsc{II}]$\lambda 6717$/[S\textsc{II}]$\lambda 6731$ ratio, or assumed to be equal to 
150 \textrm{cm} $^{-3}$ in the cases in which no [SII] line data exist, and 
the temperature in the [O\textsc{II}] zone was calculated using the 
models given in \citet{epm03}.

\begin{figure}

\includegraphics[scale=0.5]{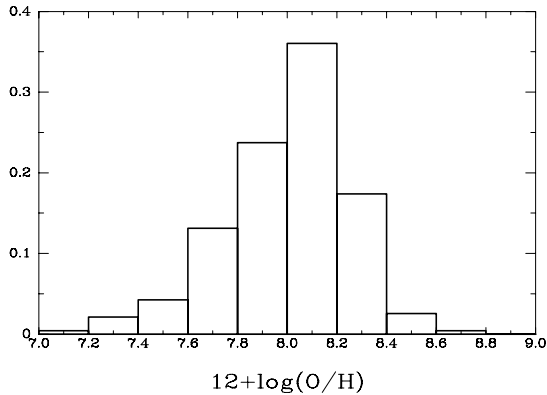}

\caption{\small \sl Metallicity distribution. Objects \emph{with}
[O\textsc{III}]$\lambda 4363$. 236 objects included.}

\label{metl_d}

\end{figure}

The oxygen abundance distribution derived for subsample Sub1 is presented in the histogram
in figure \ref{metl_d}. It is very similar to the oxygen content histogram
presented in \citet{rjt91} both in average value and scatter.

\subsection{Metallicity analysis for subsample Sub2 objects.}
For subsample Sub2, the procedure to estimate the metallicity is more 
complicated.

\subsubsection{Objects with nitrogen measurements.}
It was possible to derive oxygen abundances for 81 out of the 214 sources using
the N2 calibration introduced in \citet{denic02}. 

\[ 12+ \log (O/H) = 9.12 \pm 0.05 + 0.73 \pm 0.10 \times \log N2 \]

\noindent
where $N2$ is the [N\textsc{II}$\lambda 6584 $]/$H\alpha$ ratio.
\indent

It was shown in
\citet{epm_diaz_arbitro} that, in the case of H\textsc{II}
galaxies, this calibration produces good results.
The average metallicity of this set of 81 sources is 8.50 dex, and the
\textit{rms} is 0.27 dex. The minimum oxygen abundance is 7.8 dex. This group of 
galaxies is somewhat biased towards higher oxygen abundances with
respect to Sub1 sources.


\subsubsection{S/N approach and the use of the \citet{py00} calibration.}

We have also used another independent approach to derive the oxygen content of
Sub2 sources. First, we have used the \textit{signal-to-noise} ratio
of each spectrum, when available, to find an
upper limit to the [O\textsc{III}]$\lambda 4363$ line strength. This allows us to 
impose an upper limit on the electron temperature and hence a lower limit to 
the oxygen abundance. Only the references 6, 15, 18 and 19 provide the necessary 
information to carry out this first task. This set of objects consists on 210 objects.
Once such lower limits were calculated, those objects for which the lower
limit to $12+\log \mathrm{O/H}$ 
was greater than 8.15 dex (104 objects) were selected and their oxygen abundance was derived using the 
calibration for the \emph{upper} branch of the $R_{23}-12+\log\mathrm{(O/H})$ 
by \citet{py00}.\footnote{For 14 objects the method used to derive the
\textit{signal-to-noise} ratio yielded \emph{negative} upper limits to the
electron temperature. In these cases, the undetection of the auroral line
might be ascribed to other issues not related to the
\textit{signal-to-noise} ratio like the absorption wings of $H_{\gamma}$ or
spectral resolution effects.
These sources were excluded from the analysis.}


The value of $12+\log \mathrm{O/H}$ derived in this way turns out to be
larger than 8.15 only in 59 cases. There are however 33 objects whose metallicities fall in 
the range from 7.95 to 8.15, within the statistical error of the calibration used. These 
metallicities are then still compatible with our assumption of 
$12+\log (O/H) \geq 8.15$. The remaining 12 objects for which the derived metal content is
lower than 7.95 were excluded from the analysis. There are therefore 92 sources
for which the calibration from \citet{py00} gives potentially reliable results.
This set of 92 sources has 17 objects in common with the 81 sources previously mentioned
with metallicities derived from the [N\textsc{II}]$\lambda 6584$ line. For these 17 objects both
determinations of the oxygen abundance agree in the mean (the mean value of
the difference is 0.01dex), although the scatter is rather large
(\textit{rms}=0.25).The residuals are found not to depend on the ionization ratio.

\begin{figure}
\includegraphics[scale=0.5]{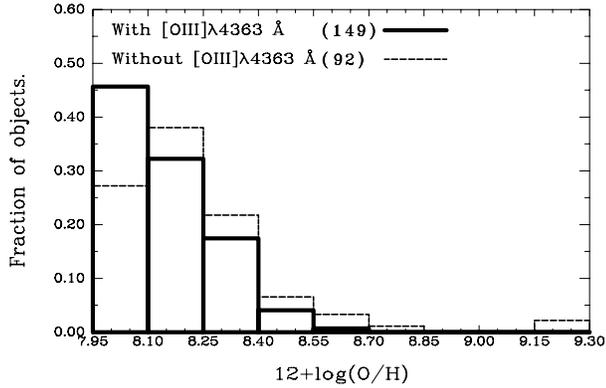}
\caption{\small \sl Metallicity distribution. The oxygen abundance was
derived using the Pilyugin (2000) calibration for the objects without 
[OIII]$\lambda 4363$. The oxygen abundance was derived using the direct method
for the subsample with [OIII]$\lambda 4363$. The difference in the mean metallicities is 0.1dex.}
\label{histo_te_pilup_6.15.18.19_d}
\end{figure}

The metallicity distribution for this set of 92 objects is compared to that of
Sub1 galaxies in figure \ref{histo_te_pilup_6.15.18.19_d}. 
%
It can be seen that both distributions overlap and show high power at 
around the same metallicities. The biggest
difference arises around $12+\log\mathrm{(O/H)}=8.00$, at which there is
almost a factor of two difference in the fraction of objects, but otherwise the
distributions are similar. It is also interesting to note that there are
objects with [O\textsc{III}]$\lambda4363$ at all oxygen abundances 
traced by objects without the line, even at fairly high oxygen abundances 
like 8.6 dex. For relatively metal-rich objects
($12+\log\mathrm{(O/H)}\geq7.95$),
those without the auroral line are only around 0.1 dex more metal rich than
objects showing the line, according to the calibration used.

\begin{figure}
\includegraphics[scale=0.5]{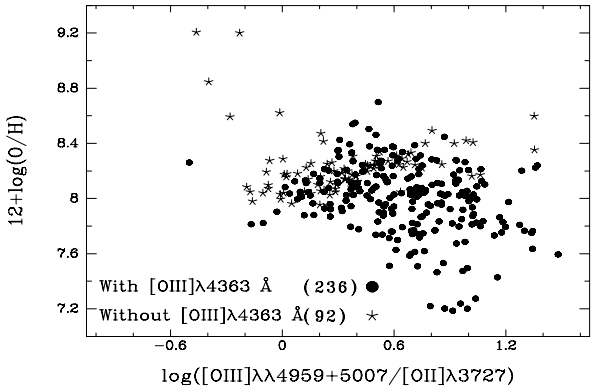}
\caption{\small \sl Metallicity against  [OIII] $ \lambda \lambda 4959+5007 $/[OII]$ \lambda 3727$ ratio.
For Sub1 objects, the direct oxygen abundance is presented. For Sub2 
objects the S/N+Pilyugin (2000) method was used. }

\label{oh.o3_o2_r1}
\end{figure}

In order to investigate to what extent this method is
affected by the differences in ionization degree among our objects, we have examined
the dependence of the derived oxygen abundances 
on the [OIII]$\lambda\lambda 4959,5007$/[OII]$\lambda 3727$ ratio. This is 
shown in figure \ref{oh.o3_o2_r1}. It is seen that at low ionization
([O\textsc{II}$\lambda4959+5007$]/[O\textsc{II}]$\lambda3727$ less than 1.0)
the metallicity scatter for Sub2 objects is rather large. Some galaxies
have a derived metallicity of around 9.2dex, and the lowest metallicities
derived for Sub2 galaxies are also found in this regime. It is also interesting to note
that, at high ionizations, it is possible to find Sub1 sources in a quite large range
of oxygen abundances.

\begin{figure}

\includegraphics[scale=0.5]{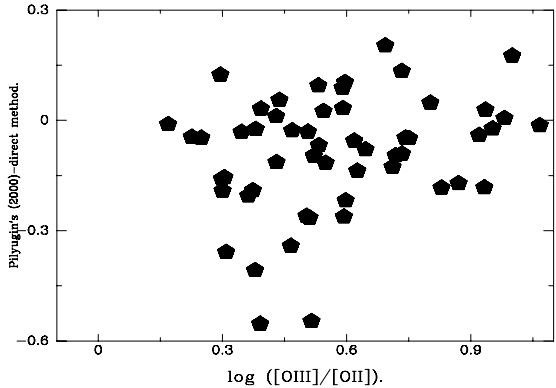}
\caption{\small \sl Metallicity residuals \textsl{vs.} 
$\log\mathrm{[OIII]/[OII]}$. Only sources from the Sub1 subsample are
plotted. 58 galaxies presented. It is seen that Pilyugin (2000) calibration
has underestimated the oxygen content of these sources, specially at 
low ionizations. At higher ionizations the residuals are closer to zero.}
\label{Oab_res_T.pilup}

\end{figure}

As a sanity check to test this method, it is applied to Sub1 sources.
Figure \ref{Oab_res_T.pilup} shows the metallicity
residuals ($\log(O/H)_{\rmn{Pilyugin.}}-\log(O/H)_{\rmn{direct.}}$) plotted 
against $\log \mathrm{[OIII]/[OII]}$. It can be seen that the S/N+\citet{py00} 
method slightly underestimates the oxygen abundances at low ionization. At higher 
values of $\log \mathrm{[OIII]/[OII]}$ the residuals are closer to zero. We 
conclude that, for most sources, the use of the upper branch of the \citet{py00} 
calibration provides reliable results, perhaps underestimating the metallicity 
for low ionization objects by around 0.2dex, even less for high-ionization objects.



We now turn to examine \citet{py00} algorithm 
more closely. This method is based in a concept which may be named ``metallicity
equivalence class''. For the upper branch, this means that objects with 
equal ionization and equal $R_{2}=$[O\textsc{II}]$\lambda3727$/H$\beta$ have equal
oxygen abundances.

According to \citet{py00} and \citet{py01}, the following quantities are defined:
$R_{2}=$[OII]$\lambda3727$/H$\beta$, $X_{2}=\log R_{2}$;
$R_{3}=$[OIII]$\lambda\lambda 4959+5007$/H$\beta$, $X_{3}=\log R_{3}$;
$R_{23}=R_{2}+R_{3}$
$X_{23}=\log R_{23}$;
$p_{2}=X_{2}-X_{23}$ and $p_{3}=X_{3}-X_{23}$.

\begin{figure}
\includegraphics[scale=0.5]{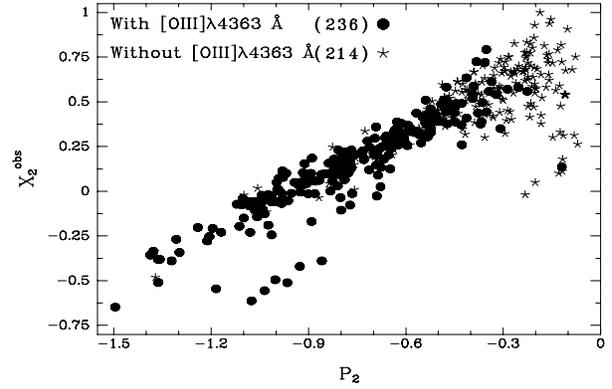}
\caption{\small \sl $X_{2}$ \textit{vs.} $p_{2}$. All objects from both subsamples.}
\label{x2.p2_all}
\end{figure}

\begin{figure}
\includegraphics[scale=0.5]{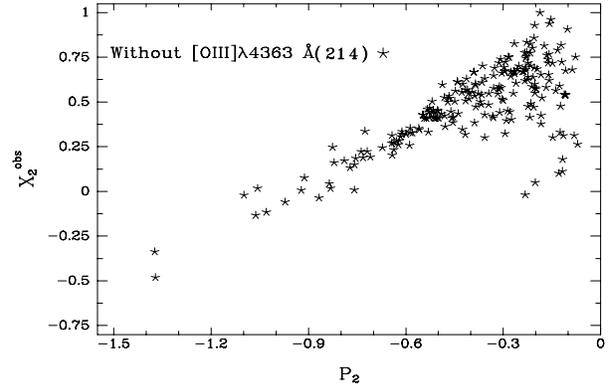}
\caption{\small \sl $X_{2}$ \textit{vs.} $p_{2}$. Subsample Sub2 objects only.}
\label{x2.p2_sub2}
\end{figure}

Figures \ref{x2.p2_all} and \ref{x2.p2_sub2} present the $X_{2}$ \textsl{vs.} $p_{2}$ 
for both subsamples, and only for subsample Sub2 respectively.
Objects belonging to the same ``metallicity equivalence class'' lie on the
same straight line. The metallicity depends on the intersect at $p_{2}=0$
of these lines.
It can be seen that very few Sub1 objects lie at values $p_{2}\geq-0.3$, while
many Sub2 objects occupy that area. This suggests that \citet{py00} method may 
not be adequate for low-ionization sources, since the \citet{py00} calibration was 
constructed using objects for which it was possible to derive oxygen abundances 
by direct methods.

In the low-ionization limit, the upper branch of the \citet{py00} calibration becomes:

\[ 12+\log \mathrm{O/H}=9.54-1.68\log R_{2} - 0.55 \mathrm{[OIII]/[OII]} \]

In this expression, when the [OIII]/[OII] ratio is very low, the
calibration becomes sensitive to $R_{2}$ only. The large scatter in $X_{2}$ observed in
figure \ref{x2.p2_sub2} for low ionization ($p_{2} \geq -0.3$)
Sub2 objects will therefore introduce a large scatter in oxygen content for these
sources which is probably unphysical since it is not observed in
low-ionization sources from the first subsample in figure \ref{oh.o3_o2_r1}.

\begin{figure}
\includegraphics[scale=0.5]{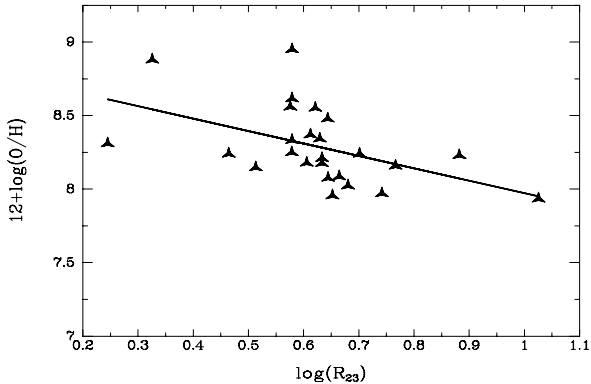}
\caption{\small \sl $12+\log\mathrm{(O/H)}$-$\log R_{23}$ calibration for
low ionization objects. This empirical calibration is used for low
ionization sources from the second subsample.}
\label{r23.o_ab_m3}
\end{figure}

It is then neccessary to re-calibrate the $12+\log\mathrm{(O/H)}-R_{23}$
relation for objects of very low ionization degree. In order to do this, we have
used a sample of low-ionization objects for which the oxygen abundance can be
determined using direct methods. This sample has been selected from
\citet{epm_diaz_arbitro} and is given in table \ref{m3_pres}.
The calibration is shown in figure \ref{r23.o_ab_m3}.

\begin{flushleft}
\begin{table*}
\caption{\small \sl List of galaxies used to derive the low-ionization
calibration.}
\begin{tabular}[t]{|cc|ccccc||}

\hline

Object. 	&  ref	&  $X_{23}$ &  $\log$([OIII]/[OII]) &  (O/H) &  $N_{e}$(cm$^{-3}$)  \\ \hline

N604C    & \cite{diaz87}  & 0.58     & -0.28    & 8.33    & 122 \\
N604E    & \cite{diaz87}  & 0.66     & -0.10    & 8.09    & 50  \\
N3310B   & \cite{pasto93} & 0.74     & -0.11    & 7.97    & 221 \\
N3310E   & \cite{pasto93} & 0.77     & -0.06    & 8.16    & 177 \\
VS38     & \cite{gar97}   & 0.51     & -0.16    & 8.15    & 129 \\
VS35     & \cite{gar97}   & 0.63     & -0.13    & 8.21    & 61  \\
VS44     & \cite{gar97}   & 0.68     & -0.15    & 8.02    & 181 \\
VS41     & \cite{gar97}   & 0.58     & -0.04    & 8.25    & 50  \\
VS24     & \cite{gar97}   & 0.62     & -0.14    & 8.55    & 79  \\
VS21     & \cite{gar97}   & 0.63     & -0.66    & 8.34    & 50  \\
VS3      & \cite{gar97}   & 0.64     & -0.02    & 8.48    & 67  \\
N79E     & \cite{denne83} & 0.64     & -0.19    & 8.08    & 50  \\
H40      & \cite{rayo82}     & 0.58     & -0.39    & 8.62    & 115 \\
NGC595   & \cite{vilchez88}     & 0.61     & -0.05    & 8.18    & 50  \\
NGC7714  & \cite{french80}    & 0.65     & -0.015   & 8.00    & 752 \\
NGC3690  & \cite{french80}    & 0.63     & -0.19    & 8.18    & 137 \\
HS1610+4539 & \cite{popes00}  & 1.03     & -0.03    & 7.93    & 50 \\
Searle5  & \cite{kr94}     & 0.33     & -0.84    & 8.88    & 50 \\
H13      & \cite{cdt02}    & 0.70     & -0.16    & 8.24    & 80 \\
H3       & \cite{cdt02}    & 0.88     & -0.41    & 8.23    & 50 \\
H4       & \cite{cdt02}    & 0.24     & -0.23    & 8.31    & 50 \\
H5       & \cite{cdt02}    & 0.46     & -0.51    & 8.24    & 50 \\
CDT1     & \cite{cdt02}    & 0.58     & -0.50    & 8.95    & 130 \\
CDT3     & \cite{cdt02}    & 0.58     & -0.55    & 8.56    & 223 \\
CDT4     & \cite{cdt02}    & 0.61     & -0.34    & 8.37    & 118 \\  \hline

\end{tabular}
\label{m3_pres}
\end{table*}

\end{flushleft}

A simple linear fit $12+\log\mathrm{(O/H)}$-$\log R_{23}$ gives:

\[ 12+\log\mathrm{(O/H)}=(8.82\pm0.20)-(0.845\pm0.31) \times \log R_{23}  \]

\noindent
whose \textit{rms} is 0.3 dex.
\indent
This new calibration allows to rederive the oxygen abundance for the
Sub2 sources of very low ionization.

\begin{figure}
\includegraphics[scale=0.5]{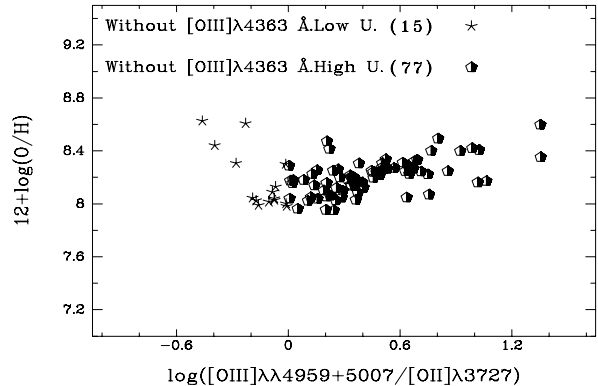}
\caption{\small \sl Metallicity \textit{vs.}  [OIII]$\lambda
\lambda4959+5007$/[OII]$\lambda 3727$ ratio. Pilyugin's 2000 and
low-ionization empirical calibration used.}
\label{oh.o3_o2_r_bis}
\end{figure}

This change in the metallicity derivation for low-ionization objects represents an
improvement as can be seen in figure \ref{oh.o3_o2_r_bis} in which the oxygen content of
low-ionization sources were calculated using the empirical calibration
given above. The large metallicity scatter of low-ionization sources from subsample 2 has been greatly reduced.
The behaviour observed in figure \ref{oh.o3_o2_r_bis} is compatible
with the evolutionary models and data presented in \citet{stasi01}. The galaxies presented
in figure \ref{oh.o3_o2_r_bis} overlap with the bulk of the data points presented in that work.

It was also mentioned above that figure \ref{oh.o3_o2_r1} shows that, in the 
case of objects from the subsample with [O\textsc{III}]$\lambda 4363$, the high ionization objects
are observed at all metallicities; on the other hand, at 
low ionizations only objects with average oxygen abundances are seen. In 
the case of the 92 objects from the second subsample, there is also a positive correlation between 
oxygen content and ionization degree.
The existence of this positive correlation was ascertained
using Spearman's test on the data presented in figure \ref{oh.o3_o2_r_bis}. To all 
practical purposes the test indicated the existence of a fairly strong 
($\rho=0.44$) positive correlation. This
correlation is worrying since higher metallicity nebulae
should show \emph{lower} ionization degrees unless the ionizing radiation is
unusually hard. However, it is seen in figure
\ref{Oab_res_T.pilup} that the upper branch of the \citet{py00} calibration
understimates the oxygen content for sources of lower ionization degree.
This means that sources with lower ionization ratios 
($\log$[OIII]$\lambda\lambda4959+5007$/[OII]$\lambda 3727 \leq 0.6$)
probably have higher metallicities than those derived from the \citet{py00}
calibration, suggesting that this unphysical correlation is an artifact 
introduced by the calibration itself. This also indicates that objects
with both high oxygen abundance and high ionization degrees do exist, since
\citet{py00} calibration becomes better at high ionization ratios. These
sources are likely to be powered by very hot radiation sources.

\begin{figure}
\includegraphics[scale=0.5]{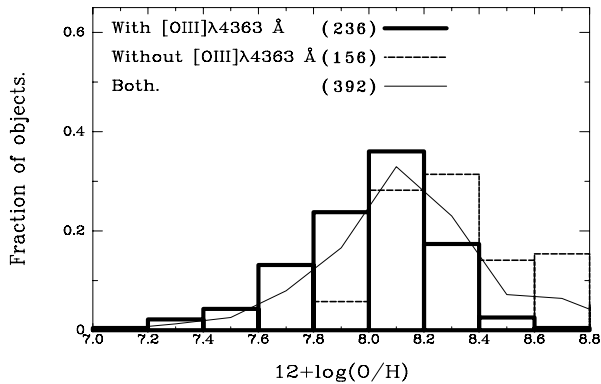}
\caption{\small \sl Metallicity distributions for both subsamples. Full 
metallicity range shown. For Sub1 objects, the oxygen abundance was 
derived using the direct method. For Sub2 objects, oxygen abundances were derived
from the empirical calibrations of 
Denicol\'o et al. (2002)(81 objects), Pilyugin (2000)(68 objects) or the
low-ionization calibration given in figure \ref{r23.o_ab_m3}(7 objects).
}
\label{te.pilup_d_full}
\end{figure}

Summarizing, using the $N2$ calibration for the objects from subsample Sub2 with the 
[N\textsc{II}]$\lambda 6584$ line available(81 objects), the empirical calibration for 
low-ionization objects
presented above (7 galaxies) and the \citet{py00} calibration for the upper 
branch (68 objects) of the $12+\log\mathrm{(O/H)}$-$\log R_{23}$ 
relation for high-ionization sources, the new metallicity distribution for
the subsample without [O\textsc{III}]$\lambda 4363$ can be compared to the
oxygen abundance distribution for the first subsample. This comparison is
presented in figure \ref{te.pilup_d_full}. 

\begin{figure}
\includegraphics[scale=0.5]{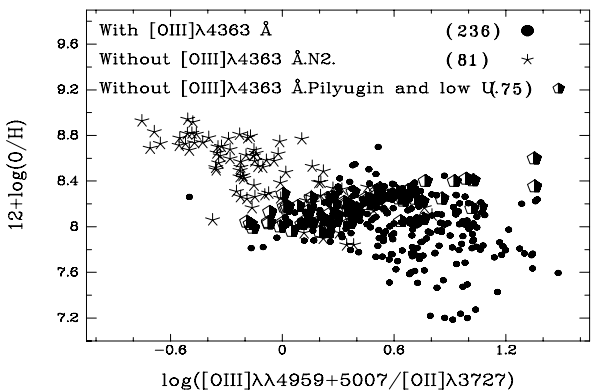}
\caption{\small \sl Metallicity \textsl{vs.} ionization ratio. All sources.}
\label{oh.o3_o2_rlast}
\end{figure}

Figures \ref{te.pilup_d_full} 
show that the oxygen abundance
distributions for both subsamples do differ. Objects without the auroral line are, on average,
\emph{at least} 0.4 dex more metal rich than objects with it.
Therefore, it is concluded that objects without the [O\textsc{III}]$\lambda 4363$ line are likely
to be of higher metalicity than objects with it. In addition, very low ionization objects are only found among
objects which do not show [OIII]$\lambda$4363.
Finally, figure \ref{oh.o3_o2_rlast} shows the relation between metallicity
and ionization ratio for all objects for which an oxygen abundance has been
derived. As expected, the global correlation is negative, and both
subsamples create a well-defined sequence. The objects for which the oxygen
abundance was derived either from the calibration presented above or the
upper branch of the \citet{py00} calibration lie on the region shared by the
first subsample and the objects with [N\textsc{II}]$\lambda 6584$ measurements from the
second subsample. The calibration used makes their oxygen abundances to be
those of objects with [O\textsc{III}]$\lambda 4363$ of similar ionization
ratio. 
However, figure \ref{Oab_res_T.pilup} shows that for low ionization
sources with [O\textsc{III}]$\lambda 4363$, the use of the \cite{py00} calibration
underestimates the oxygen abundance by 0.15dex. This opens the possibility that the
oxygen content of low-ionization subsample Sub2 sources whose metallicity was derived using
\cite{py00} expressions or the low-ionization calibration introduced here 
might be underestimated. If this turns out to be the case, and some fraction 
of these low-ionization sources actually lies closer to the other members of 
the second subsample with nitrogen measurements in figure \ref{oh.o3_o2_rlast}, the
separation in oxygen abundance between subsamples Sub1 and Sub2 would be somewhat larger.

\section{The LCBG-like subsample.}

Luminous Compact Blue Galaxies, hereafter LCBGs, are the high luminosity end
of BCGs. 
They are operationally defined as luminous 
(M$_{B}$ more luminous than -17.5), blue ($B-V$ bluer than 0.6) and
compact $\mu_{B}\leq21.5$\textrm{mag arcsec}$^{-2}$ systems.
 Their spectra indicate that they are undergoing a major starburst,
which produces a significant fraction of their light output. This starburst
enhances their surface brightnesses, making it possible to see them at large
distances. Hubble Space Telescope images of LCBG show the presence of an important underlying stellar
population too. Spectroscopic studies of LCBG can be found in \citet{guz97},
\citet{phi97} and \citet{hoyos2004}. HST imaging is presented and
discussed in \citet{koo94} and \citet{guz98}.
In \citet{guz97}, LCBGs are divided into H\textsc{II}-like and
Nuclear Starburst-like types. The work presented in \citet{hoyos2004} further
highlitghts the similarities between LCBGs and the most luminous H\textsc{II}
galaxies.

\subsection{Definition of LCBG-like H\textsc{II} galaxies.}

At least some fraction of the population of intermediate-redshift LCBGs can be 
considered to be very similar to bright, local H\textsc{II} galaxies (see again \citealt{guz97},
\citealt{phi97} and \citealt{hoyos2004}). We here define LCBG-like H\textsc{II} galaxies
as the subsample of local H\textsc{II} galaxies whose preperties resemble those of higher-redshift LCBGs.

In order to extract such subsample of local H\textsc{II} galaxies with properties similar to
higher-redshift LCBGs, the first step is to find galaxies more luminous than 
$M_{B}=-17.5$ in the sample studied here. The approach we adopt is to represent 
the observed blue absolute magnitude ($M_{B}$) versus the estimator of the blue absolute 
magnitude presented in \citet{jmelnickxx} ($B_{c}$).

\[ B_{c}=79.4-2.5 \log \frac{L(H\beta)\hbox{ergs}^{-1}}{W_{\beta}\hbox{(\AA)}} \]

This calibration is introduced to take into account possible aperture/distance or line
contamination effects. The number $B_{c}$ only probes the continuum strength of the fraction of the 
galaxy that fell inside the slit, but the blue absolute magnitude is sensitive to all light 
within the passband. This calibration tries to correct for these effects with the aim of deriving an
estimate of the blue absolute magnitude for the galaxies we are studying.

\begin{figure}
\includegraphics[scale=0.5]{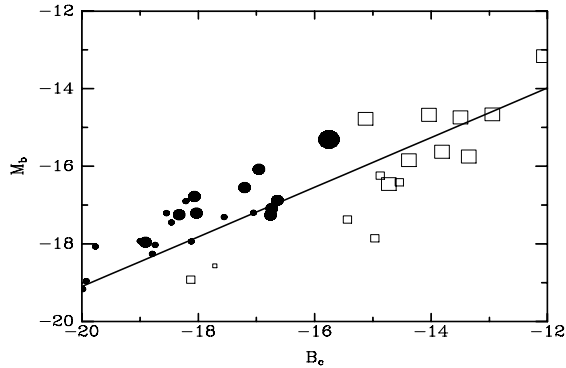}
\caption{\small \sl $M_{B}$ vs $B_{c}$ calibration. Black, solid dots represent 
objects from reference 19. Open squares are objects from references 
13, 16, 17 and 18. Small symbols: galaxies further than $cz=3600$km s$^{-1}$.
medium symbols: galaxies in the range $1800$km s$^{-1}$ $\leq cz \leq 3600$ km s$^{-1}$.
Big symbols:galaxies nearer than 1800 km s$^{-1}$. }
\label{bc_calib}
\end{figure}

The calibration is shown in figure \ref{bc_calib}. This plot
presents $M_{B}$ \textsl{vs.} $B_{c}$ for the galaxies from the references
13, 16, 17, 18 and 19 for which $M_{B}$ was available. In the case of galaxies
from the Hamburg Quasar Survey, $M_{B}$ was reported in \citet{popes00}. In the case of galaxies
from the First and Second Byurakan Surveys, $M_{B}$ was found in the NED or HyperLeda databases.
The straight line shown is a least-squares fit to the calibration. The fit expression is:

\[ M_{B}=0.64 \pm 0.06 B_{c}-6.3\pm1.0 \]

Its scatter is around 1.0 magnitudes, and the residuals are found not to depend on 
$\log (\mathrm{Slit Width}\times cz)$. 
In figure \ref{bc_calib}, $B_{c}$ was calculated using spectra with slit widths of 4\arcsec \citep{popes00}
and 2\arcsec (objects from the Byurakan surveys). Most spectra in this work were taken with 
apertures between these two sizes, so the above calibration can be applied for them. Aperture 
effects for spectra taken with larger apertures or of more distant objects will likely be 
less important. According to this calibration, objects belonging to the LCBG-like subsample
are required to have $B_{c}$ lower than -17.5, which is the limit adopted.

The color requirement was not straightforward to apply, because not many
$B-V$ colors for the H\textsc{II} galaxies included in the sample were
available in the literature. For this reason, a simple alternative method,
based on spectroscopic criteria had to be developed. 

The galaxy sample presented in \citet{cairos_01_a} and \citet{cairos_01_b}
provides with $B-V$ colors for a sample of 28 blue compact galaxies, six of
them appear in the \citet{rjt91} catalogue. This smaller sample is shown in
table \ref{m4_pres}.

\begin{flushleft}
\begin{table*}
\caption{\small \sl Galaxies used to re-define the color-criterium.
Notes:
(a) For Mrk370 and IIZw71 this number is defined as $1.3\times W_{5007}$;
(b) This is the reddening-corrected $I_{\lambda}/I_{\beta}$ ratio;
(1) Given values are $H\beta$weighted averages of the different zones defined in the
respective papers;
(2) Integrated spectrophotometry.}
\begin{tabular}[t]{|llllllll|}

\hline

Object       &   B-V   &  ref            & [O\textsc{II}]$\lambda3727$(b) & 
[O\textsc{III}]$\lambda5007$(b) & $W_{3727}$ & $W_{\beta}$ & $W_{4959+5007}$(a) \\ \hline
Tol0127-397  &  0.56   & \cite{rjt91}    & 3.278  & 2.164  &  73   & 38 &   130  \\ 
UM417        &  0.36   & \cite{rjt91}    & 0.330  & 5.566  &  20   & 144 &  903  \\
Mrk370(1)    &  0.48   & \cite{cairos02} & 2.803  & 1.678  &  60.7 & 17.0 & 37.5 \\
IIZw40       &  0.52   & \cite{rjt91}    & 0.440  & 7.76   &  79   & 268  & 2122 \\
Mrk36        &  0.39   & \cite{rjt91}    & 0.988  & 5.506  &  42   & 70   & 432  \\
UM462(1)     &  0.47   & \cite{rjt91}    & 1.777  & 5.320  &  96.1 & 102  & 615.8 \\
IIZw71(2)    &  0.55   & \cite{jansen00} & 4.397  & 2.49   &  33.2 & 7.5 &  22.53 \\ \hline

\end{tabular}
\label{m4_pres}
\end{table*}

\end{flushleft}

The adopted approach is to derive a least-squares fit to the
observed colors as a function of the continuum strength ratio
$x=\frac{\hbox{[OIII]}\lambda 5007 \times W_{3727}}{\hbox{[OII]}\lambda 3727\times W_{4959+5007}}$.
The fit is:

\[  B-V=(0.54\pm0.04)+(0.35\pm0.16) \log x \]

Using this expression, the color condition is translated into $1.57\leq x$,
which is the condition applied. 

Unfortunately, not all references report the observed equivalent widths of
the two oxygen lines required. Again, only the \citet{rjt91} catalogue
provides all the information.

However, all the objects from \citet{rjt91} which show 
[O\textsc{III}]$\lambda4363$ included in the presented sample meet 
this requirement, and 96\% of the objects without the auroral line from the 
\citet{rjt91} sample also match the condition. Therefore, it is assumed that
\emph{all} the H\textsc{II} galaxies studied satisfy this condition.

This is not surprising, since the objects presented here were selected
from objective prism surveys searching for either strong emission lines
or UV excesses. The Tololo and UM surveys, looking for strong lined objects will
pick up very blue objects, or compact starbursts with a weak continuum due to the 
presence of massive, young blue stars. The Markarian or Byurakan surveys, searching for
galaxies presenting a UV excess naturally select blue objects. In addition, in these 
surveys, the photographic plates were more blue sensitive.

Unfortunately, no constraint on the surface brightness or half-light radius
can be used with the data at hand. Galaxies satisfying the first two criteria
are said to belong to the LCBG-like sample.

In total, 50 objects from subsample Sub1 and 117 sources from the
subsample Sub2 were selected. These 167 objects are
considered to be LCBG-like H\textsc{II} galaxies.

\subsection{Properties of LCBG-like objects.}

\begin{figure}
\includegraphics[scale=0.5]{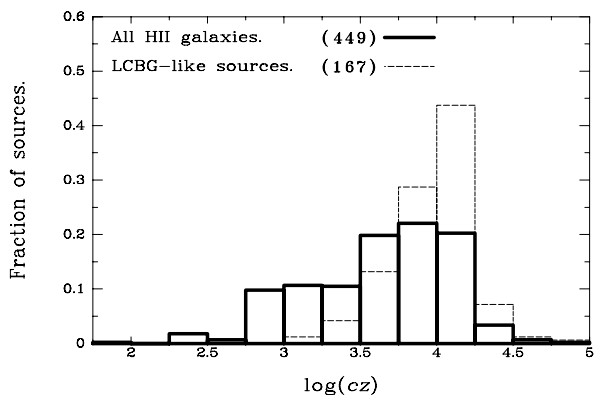}
\caption{\small \sl Radial velocity distributions.}
\label{logcz_d_lcbg}
\end{figure}

\begin{figure}
\includegraphics[scale=0.5]{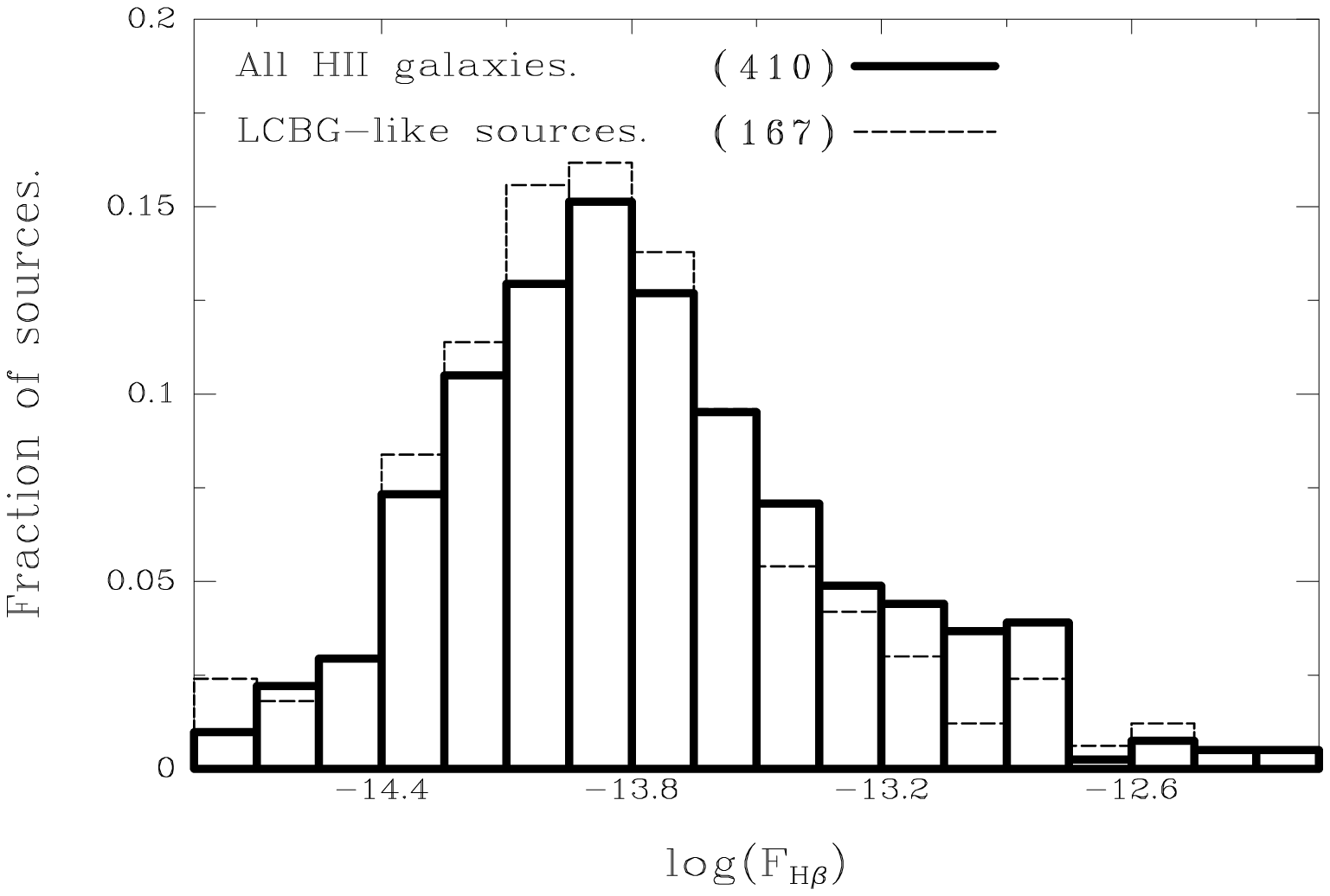}
\caption{\small \sl  Observed H$\beta$ flux distributions.}
\label{logfhb_d_lcbg}
\end{figure}

\begin{figure}
\includegraphics[scale=0.5]{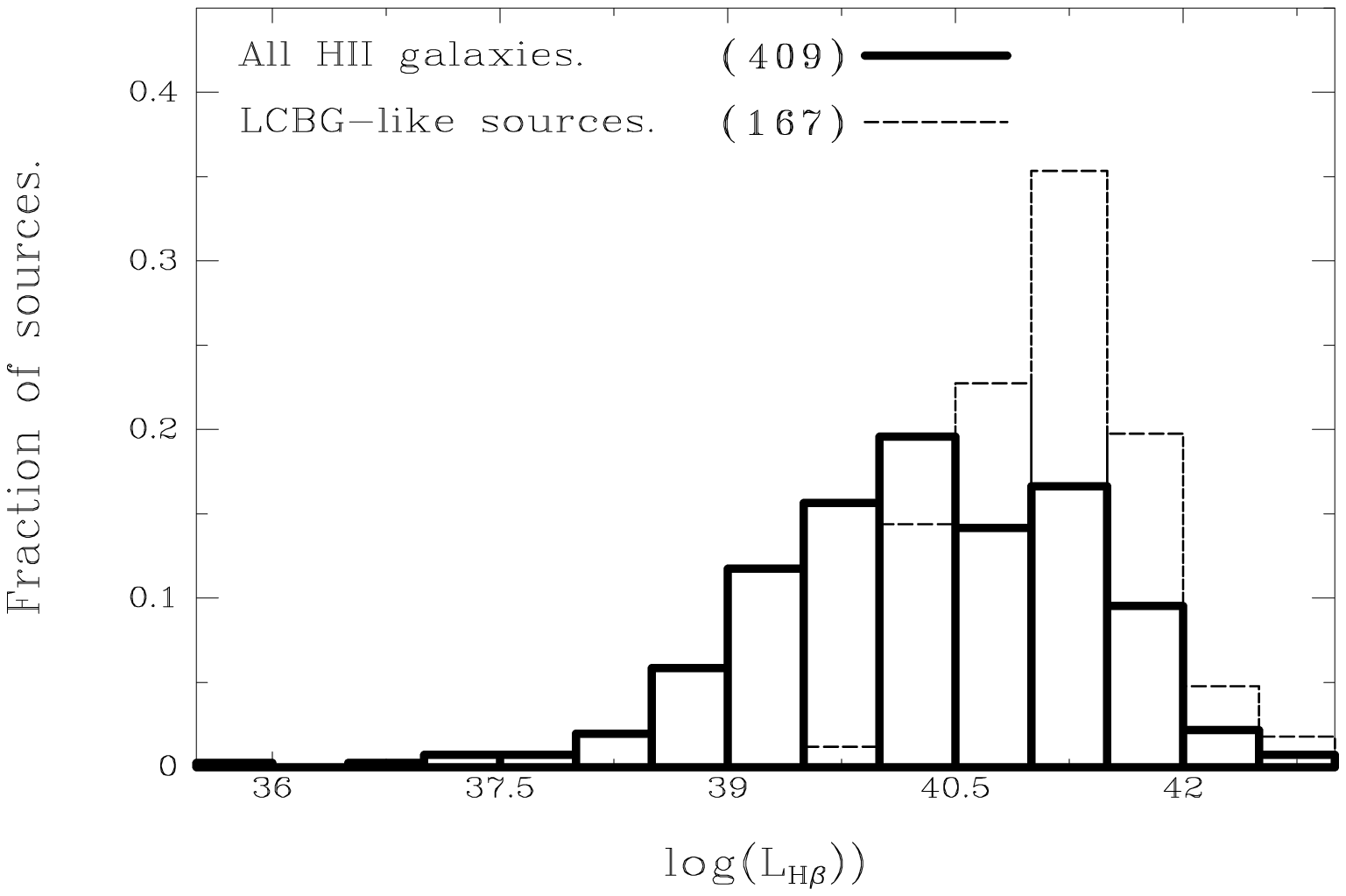}
\caption{\small \sl H$\beta$ luminosity distributions.}
\label{loglhb_d_lcbg}
\end{figure}

\begin{figure}
\includegraphics[scale=0.5]{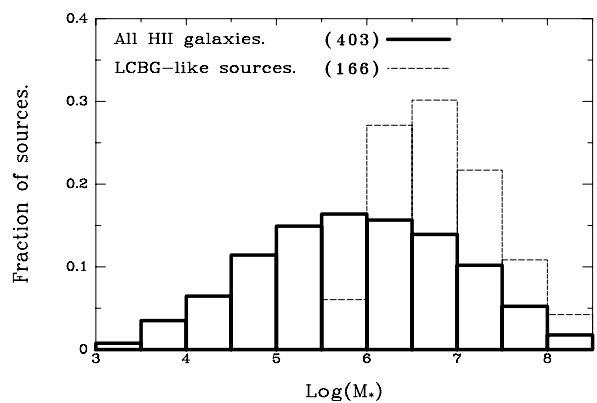}
\caption{\small \sl  Mass of the ionizing cluster distributions.}
\label{logMc_d_lcbg}
\end{figure}

\begin{figure}
\includegraphics[scale=0.5]{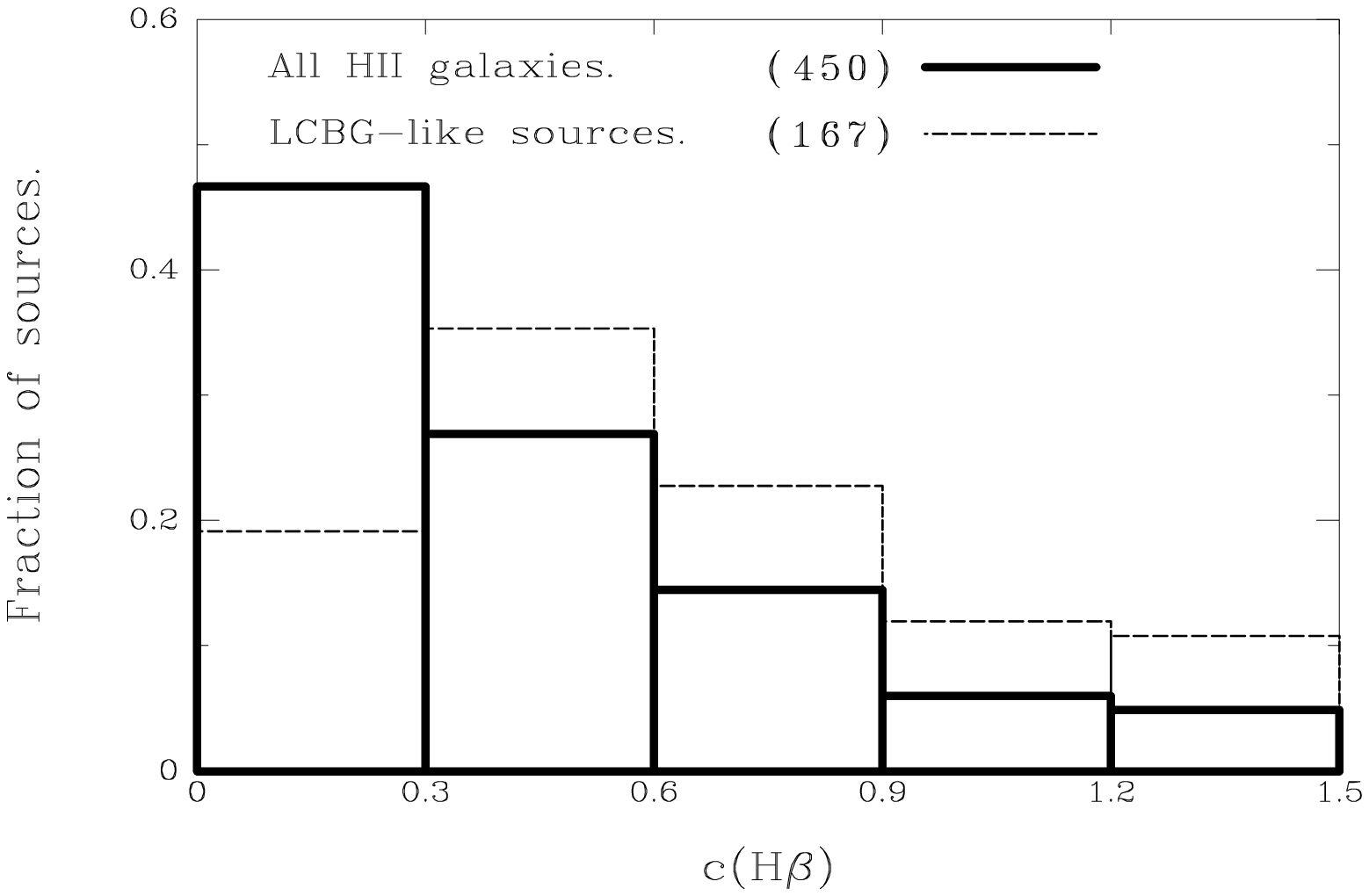}
\caption{\small \sl c(H$\beta$) distributions.}
\label{chb_d_lcbg}
\end{figure}

\begin{figure}
\includegraphics[scale=0.5]{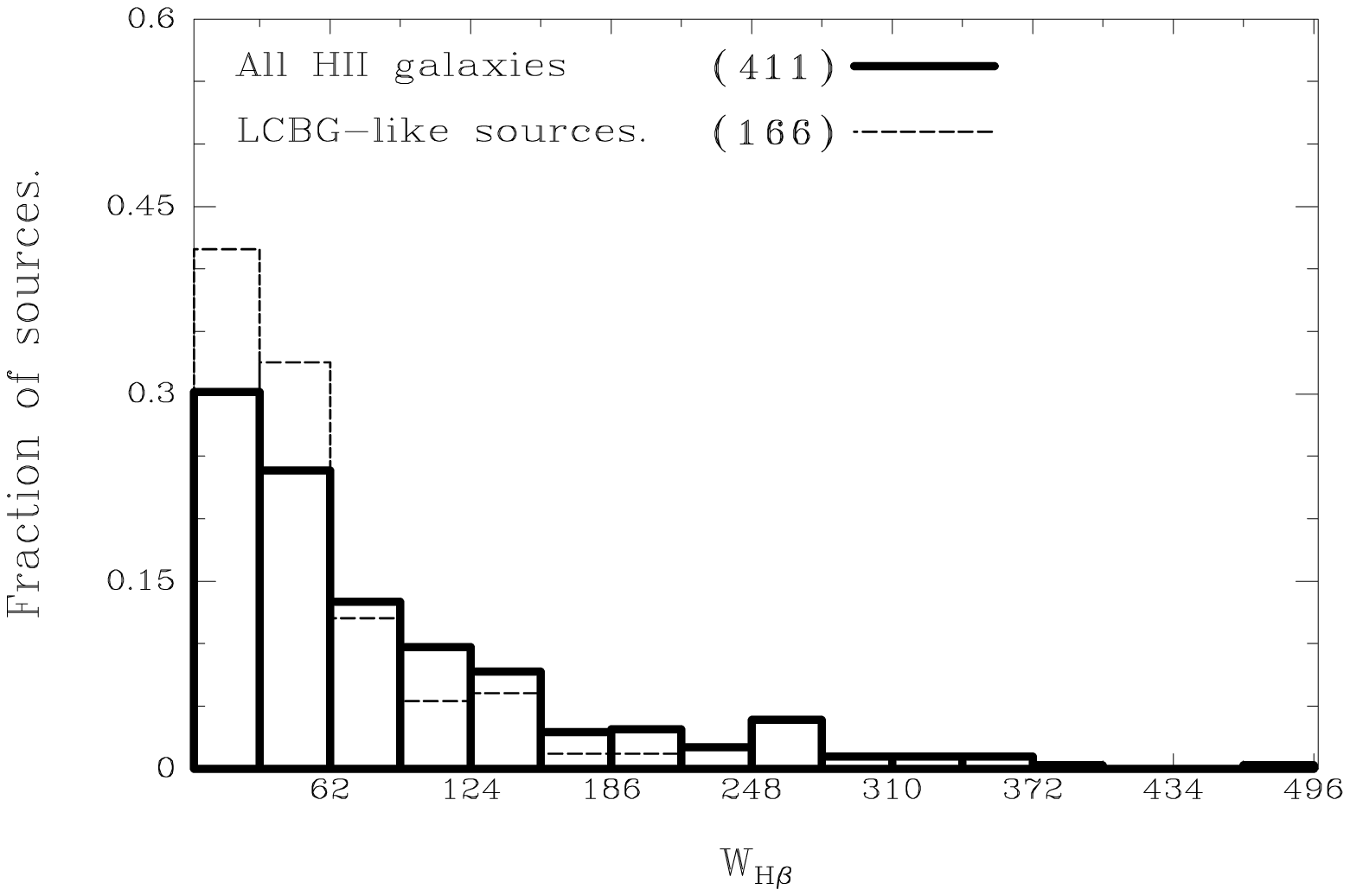}
\caption{\small \sl  W$_{\beta}$ distributions.}
\label{wb_d_lcbg}
\end{figure}

\begin{figure}
\includegraphics[scale=0.5]{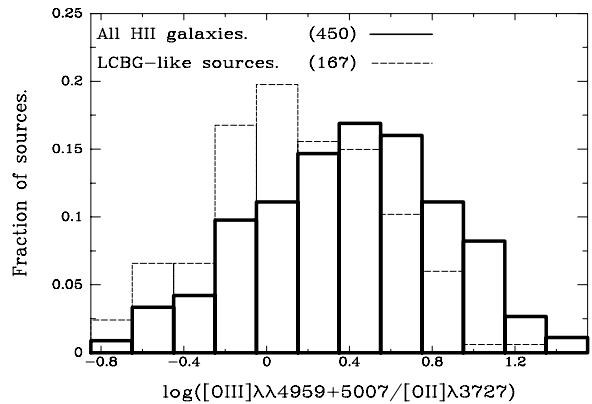}
\caption{\small \sl  Ionization ratio distributions.}
\label{logo3_o2_d_lcbg}
\end{figure}

\begin{figure}
\includegraphics[scale=0.5]{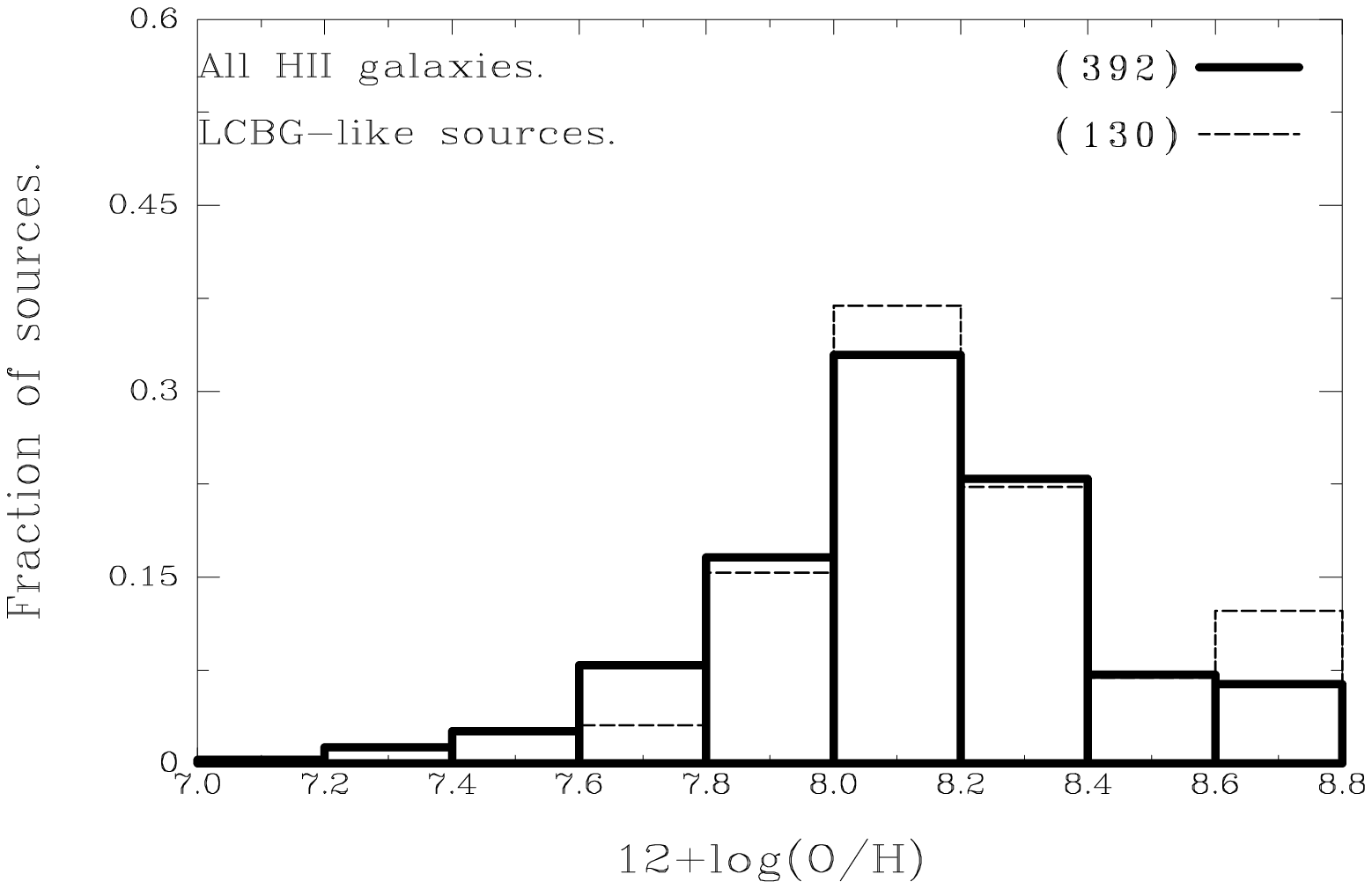}
\caption{\small \sl  Metallicity distributions.}
\label{oab_d_lcbg}
\end{figure}

The aim of this work is to see where LCBG-like sources properties fit within
the frame of H\textsc{II} galaxies in general. In order do this, the
distributions of several quantities were drawn.
Figures \ref{logcz_d_lcbg} to \ref{oab_d_lcbg} give the corresponding
distributions of the quantities.

Figure \ref{logcz_d_lcbg} indicates that LCBG-like H\textsc{II} galaxies
are mainly found at large distances. Only a tiny fraction of LCBG-like sources
is located at redshifts lower than $\sim0.01$, while this is the median value of
the redshift distribution for less luminous systems.
Figure \ref{logfhb_d_lcbg} shows that the observed H$\beta$ flux distribution 
from LCBG-like objects is remarkably similar to that of the whole 
sample.
In figure \ref{loglhb_d_lcbg} it is seen that almost all LCBG-like galaxies
have H$\beta$ luminosities greater than $10^{40}$\textrm{erg s$^{-1}$}. Only
around 20 normal H\textsc{II} galaxies out of 242 are found with line luminosities greater than 
$10^{41}$\textrm{erg s$^{-1}$}. This suggests that a line
luminosity around $3 \times 10^{40}$\textrm{erg s$^{-1}$} might be
considered as the cut-off for H\textsc{II} galaxies similar to 
LCBG. 
This is a very important difference, since it means that 
this bright subsample of H\textsc{II} galaxies is powered by a greater number of stars.
This is confirmed in figure \ref{logMc_d_lcbg}, which presents the $\log M_{*}$ distribution for both 
samples. It is seen that LCBG-like H\textsc{II} galaxies have ionizing
cluster masses greater than $10^{6}M_{\odot}$.
Only 20 lower-luminosity H\textsc{II} galaxies have clusters that massive.
However, since the calculated 
masses are only lower limits to the actual values due to
photon-escape, presence of dust and the systematic error in the
\textit{starburst} equivalent width introduced by the underlying, non-ionizing 
population (partially corrected for by the use of the upper envelope of the
W$_{\beta}$ \textit{vs.} $\log(\mbox{[OIII]/[OII]})$ presented in 
figure \ref{wb.o3_o2_r}), LCBG-like H\textsc{II} galaxies will harbor ionizing cluster 
much more massive than this limit.
This is the single, most important
difference between LCBG-like  H\textsc{II} galaxies and lower luminosity
systems. 
The extinction distribution, shown in figure \ref{chb_d_lcbg}, indicates
that there is a lack of LCBG-like sources with very low extinctions. At 
higher dust contents, c(H$\beta$)$\geq 0.30$, both distributions follow 
each other rather closely, though. This distribution also shows that
bright H\textsc{II} galaxies are not affected by uncertainties in the 
extinction to a greater extent than the rest of the galaxies studied.

Figure \ref{wb_d_lcbg} shows the existence of an upper limit to the equivalent
width of the more luminous systems of around 200\AA. The median values are
46.5\AA{} for the LCBG-like subsample and 77.5\AA{} for the rest of less luminous
systems. These numbers show that large equivalent widths are mainly found among
low luminosity H\textsc{II} galaxies and young starburst ages. Therefore,
LCBG-like H\textsc{II} galaxies have probably built larger underlying 
stellar populations. This is further supported by the fact that the 
LCBG-like histogram shows a higher occupation number at very low 
equivalent widths.
Figure \ref{logo3_o2_d_lcbg} shows that the excitation is lower in
LCBG-like H\textsc{II} galaxies. This suggests that the ionizing star clusters
of LCBG-like H\textsc{II} galaxies are probably more evolved than those of
less luminous H\textsc{II} galaxies.
The oxygen abundance of LCBG-like systems compared to the whole sample can
be seen in figure \ref{oab_d_lcbg}. This plot shows that there is a slight
bias in the metal content. LCBG-like galaxies never show oxygen abundances
lower than 7.6. At the same time, a very metal-rich object is more likely to be a
LCBG-like galaxy. However, there seems to be no relationship between
metallicity and luminosity since LCBG-like galaxies span pretty much the
same metallicity range as the bulk of the whole sample of H\textsc{II} galaxies.
In addition, figure  \ref{oab_d_lcbg} shows that the observed differences
in the distributions of the ionization ratio shown in figure \ref{logo3_o2_d_lcbg}
can't arise from a metallicity effect.

It is also enlightening to see whether LCBG-like galaxies are 
similar to H\textsc{II} galaxies with [O\textsc{III}]$\lambda4363$, or
if they resemble H\textsc{II} galaxies without the auroral line.
In order to do this, the dot product between the LCBG-like subsample
 and both Sub1 and Sub2 subsamples probability densities presented above
(i.e. the histograms) were derived. Even though the probability densities for
subsamples Sub1 and Sub2 are not orthogonal for any property studied, this
should indicate which subsample is more similar to LCBG-like galaxies.

{\tiny
\begin{flushleft}
\begin{table}
\caption{\small \sl Normalised dot products between LCBG-like subsample histograms and
subsamples Sub1 and Sub2 probability densities.}
\begin{tabular}[t]{|l|cc|} \hline
Property.         &  With   &  Without \\ \hline
B$_{\mathrm{c}}$  &  0.622  &  1.228    \\
L$H\beta$      &  0.911  &  1.085    \\
$\log$[OIII]/[OII]&  0.560  &  0.955    \\
c(H$\beta$)       &  0.446  &  0.965    \\
$W_{\beta}$       &  0.889  &  0.880    \\
$\log M_{*}$      &  0.735  &  1.123   \\
$12+\log (O/H)$   &  0.905  &  0.935   \\ \hline

\end{tabular}
\label{escalares}
\end{table}
\end{flushleft}
}

Table \ref{escalares} clearly shows that LCBG-like sources are more similar
to objects not showing the auroral line [O\textsc{III}]$\lambda 4363$. This table
indicates that if a distant LCBG does not show the [O\textsc{III}]$\lambda 4363$ line, it
is likely that this object will present a massive underlying stellar population, high metallicity 
and low ionization.

\section{Summary and Conclussions.}

We have conducted a statistical study of a very large spectroscopic sample of H\textsc{II}
galaxies from the literature. We have compared galaxies with and without the
[O\textsc{III}]$\lambda 4363$ line, and we have defined a control sample 
which can be used to investigate the nature of LCBGs at intermediate \textit{z}.

It has been found that H$\beta$ fluxes are larger for objects showing the 
[O\textsc{III}]$\lambda 4363$ line, even though the H$\beta$ luminosity 
distributions for galaxies with and without the auroral line are very similar.
This is in part because objects not showing the auroral line are more distant and 
their extinction is higher. However, it has been shown that the undetection of 
the [O\textsc{III}]$\lambda 4363$ line is a real metallicity effect for at least 
some fraction of cases. Objects without the auroral line are about 0.4dex more metal 
rich than objects from subsample Sub1. The analysis of the 
[O\textsc{III}]/[O\textsc{II}] to $12+\log (O/H)$ relationship reveals the existence
 of high-ionization, metal rich objects without the auroral line.
Objects from the second subsample are found to harbour more
massive star clusters, although the differences in excitation 
between the two subsamples indicates that subsample Sub2 sources are probably 
powered by somewhat older star clusters.

LCBG-like sources are clearly further away than the average local
H\textsc{II} galaxy. The H$\beta$ luminosities of LCBG-like systems 
are much greater. This is a very important difference between
the two subsamples. We have also found that their ionizing star clusters 
are more massive that those of lower luminosity H\textsc{II} galaxies.
LCBG-like H\textsc{II} galaxies have been found to posses larger (and hence 
probably older) underlying populations, and their ionizing star clusters 
are also more evolved and massive. LCBG-like sources are marginally more metal-rich 
than the average H\textsc{II} galaxy, but not enough to explain the observed
differences in the ionization ratio [O\textsc{III}]/[O\textsc{II}].

We have also shown that LCBG-like sources are more similar
to objects without the auroral line [O\textsc{III}]$\lambda 4363$, implying
that any local control sample designed to study high-redshift LCBGs is to be made of
galaxies not showing the [O\textsc{III}]$\lambda 4363$ line. If one observes a distant 
LCBG and is unable to detect the [O\textsc{III}]$\lambda 4363$ line, it
is likely that this object will present a massive underlying stellar population, high metallicity 
and low ionization.

\section*{Acknowledgments}

We would like to thank Dr. P\'erez-Montero for his help in
building table \ref{m3_pres} and deriving the metallicities there included. 
We would also like to thank an anonymous referee for his/her valuable comments.
We also acknowledge financial support from the DGICYT grants 
AYA-2000-0973, AYA-2004-08260-CO3-03, and from the MECD FPU grant AP2000-1389. 
This research has made extensive use of the NASA/IPAC Extragalactic Database (NED)
which is operated by the Jet Propulsion Laboratory, California Institute 
of Technology, under contract with the National Aeronautics and Space Administration. 
We have also used the HyperLeda database, which can be reached at
 http://www-obs.univ-lyon1.fr/hypercat/intro.html

\label{lastpage}

\end{document}